\documentclass[twocolumn,floatfix,showpacs,prd,aps,tightenlines]{revtex4}
\usepackage{graphicx}
\usepackage{amsmath}
\usepackage{psfrag}
\usepackage{dcolumn}% Align table columns on decimal point
\usepackage{bm}% bold math

\newcommand{\bea}{\begin{eqnarray}}
\newcommand{\eea}{\end{eqnarray}}
\newcommand{\beq}{\begin{equation}}
\newcommand{\eq}{\end{equation}}

\begin{document}

\title{Algebraic Classification of Numerical Spacetimes and 
Black-Hole-Binary Remnants}

\author {Manuela Campanelli}
\author{Carlos O. Lousto}
\author{Yosef Zlochower} 
\affiliation{Center for Computational Relativity and Gravitation, and\\
School of Mathematical Sciences,
Rochester Institute of Technology, 78 Lomb Memorial Drive, Rochester,
 New York 14623}

\date{\today}

\begin{abstract}
In this paper we develop a technique for determining the algebraic
classification of a numerically generated spacetime, possibly resulting from a
generic black-hole-binary merger, using the Newman-Penrose Weyl
scalars. We demonstrate these techniques for a test case involving a
close binary with arbitrarily oriented spins and unequal masses. We
find that, post merger, the spacetime quickly approaches Petrov type
II, and only approaches type D on much longer timescales. These
techniques, in combination with techniques for evaluating acceleration and NUT
parameters, allow us to begin to explore the validity of the ``no-hair
theorem'' for generic merging-black-hole spacetimes.
\end{abstract}

 \maketitle

\section{Introduction}
\label{sec:intro}

The recent breakthroughs in numerical
relativity~\cite{Pretorius:2005gq, Campanelli:2005dd, Baker:2005vv}
that allowed for stable evolutions of black-hole-binary spacetimes led
to many advancements in our understanding of black-hole physics, and
it is now possible to accurately simulate the merger process and
examine its effects in this highly non-linear
regime~\cite{Campanelli:2006gf, Baker:2006yw, Campanelli:2006uy,
Campanelli:2006fg, Campanelli:2006fy, Pretorius:2006tp,
Pretorius:2007jn, Baker:2006ha, Bruegmann:2006at, Buonanno:2006ui,
Baker:2006kr, Scheel:2006gg, Baker:2007fb, Marronetti:2007ya,
Pfeiffer:2007yz}.  Black-hole binaries radiate between $2\%$ and $8\%$
of their total mass and up to $40\%$ of their angular momenta in the
last few orbits, depending on the magnitude and direction of the  spin
components, during the merger~\cite{Campanelli:2006uy,
Campanelli:2006fg, Campanelli:2006fy} (ultra-relativistic head-on
black-hole mergers can radiate up to $\sim14\%$ of their
mass~\cite{Sperhake:2008ga}).  In addition, the radiation of net
linear momentum by a black-hole binary leads to the recoil of the
final remnant hole~\cite{Campanelli:2004zw,  Herrmann:2006ks,
Baker:2006vn, Sopuerta:2006wj, Gonzalez:2006md, Sopuerta:2006et,
Herrmann:2006cd, Herrmann:2007zz, Herrmann:2007ac, Campanelli:2007ew,
Koppitz:2007ev, Choi:2007eu, Gonzalez:2007hi, Baker:2007gi,
Campanelli:2007cga, Berti:2007fi, Tichy:2007hk, Herrmann:2007ex,
Brugmann:2007zj, Schnittman:2007ij, Krishnan:2007pu,
HolleyBockelmann:2007eh, Pollney:2007ss, Dain:2008ck}, which can have
astrophysically observable important effects~\cite{Redmount:1989,
Merritt:2004xa, Campanelli:2007ew, Gualandris:2007nm,
HolleyBockelmann:2007eh, Kapoor76, Bogdanovic:2007hp, Loeb:2007wz,
Bonning:2007vt, HolleyBockelmann:2007eh, Komossa:2008qd,
Komossa:2008ye, Shields:2008kn} and represents a possible strong-field
test of General Relativity (GR).

In addition to important astrophysical applications, the two body problem in
GR is intrinsically interesting because it
provides the framework for analyzing the behavior of the theory in the
highly-nonlinear, highly-dynamical, non-symmetrical regime.
For example, the cosmic censorship hypothesis, that states that
singularities in the universe should be cloaked by a horizon
is under active investigation~\cite{Campanelli:2006uy,
Campanelli:2006fg, Campanelli:2006fy, Rezzolla:2007rd,
Sperhake:2007gu}.  In this paper we are interested in verifying the
``no hair theorem", which states that all black holes eventually relax
into a state that can be described
by three parameters, the mass, spin, and charge. Hence, the
final merger remnants from multi-black-hole
mergers~\cite{Campanelli:2007ea, Lousto:2007rj} should be Kerr black
hole~\cite{Mazur:2000pn}.

The problem of determining the geometry of the final stage of a
black-hole binary merger arises as a practical question even in perturbative
techniques, such as the  Lazarus approach~\cite{Baker:2001sf,
Campanelli:2005ia}, which used a combined numerical and perturbative
approach to simulate the waveforms from a binary merger. In the
context of the Lazarus approach, it is crucial to determine when
the transition from numerical to perturbative evolutions is  possible,
i.e.\ when the full numerical simulation could be approximated by
(relatively small) perturbations of a Kerr-rotating black hole, and a
diagnostic, the S-invariant~\cite{Baker00a}
\begin{equation}\label{eq:S}
S=27J^2/I^3,
\end{equation}
that is identically $1$ for a Kerr spacetime, was developed to measure
the closeness of the spacetime to an algebraically special type II.
However, the S-invariant by itself is not sufficient to demonstrate
that the spacetime is near Kerr because it does not distinguish
between type II and type D spacetimes, nor does it imply that the
acceleration and NUT parameters vanish.

More recently, with the availability of new long term evolutions, one
of the consistency tests performed is the agreement of the total angular
momentum of the system when computed in three different ways: by
measuring the angular momentum (and mass) of the remnant black
hole~\cite{Campanelli:2006fg, Campanelli:2006fy, Krishnan:2007pu}
using the isolated horizon formulae~\cite{Dreyer02a}, by measuring the
total energy and angular momentum radiated~\cite{Campanelli99,
Lousto:2007mh} and subtracting it from the total initial values, and
by looking at the quasi-normal frequencies of the late-time waveforms
and associate them with those of a rotating Kerr hole with mass $M$
and angular momentum per mass $a$~\cite{Dain:2008ck}.  The rough
agreement of those values represents indirect evidence that the final
black hole is of the Kerr type. Furthermore, in
Ref.~\cite{Scheel:2008rj}, where the authors of that paper presented
very-high-accuracy waveforms from the merger of an equal-mass black-hole
binary, it was shown that the minimum and maximum values of the scalar
curvature on the remnant horizon agreed with the Kerr values.

No hair theorems assume a stationary Killing
vector~\cite{Mazur:2000pn} as characterizations of the Kerr
geometry~\cite{Mars:1999yn, Mars:2000gb}.  While one can classify
spacetimes based on their symmetry properties, here we will use a
classification method based on the algebraic properties of  generic
spacetimes without a-priori assumptions about symmetries.

Demonstrating that the remnant of a black-hole merger approaches Kerr
asymptotically (in time) would also help answer open questions about
the stability of Kerr  under arbitrary perturbations.  The stability
of the Kerr spacetime under linear perturbations has only been proven
mode-by-mode~\cite{Whiting:1989vc}, and the interior of the hole may
even be unstable~\cite{Dotti:2008yr}.  Hence a study of the invariant
geometrical properties of the black-hole merger, which would yield a
highly-nontrivial perturbation of the `Kerr' background, may answer
many open questions.

\section{Mathematical Techniques}
\label{sec:mathtechniques}

In the following sections we will use the convention that Latin indices
range over the spatial coordinates (i.e.\ $a=(1,2,3)$) and Greek indices
range over all four coordinates.
\subsection{Petrov type}

The Petrov classification of a generic spacetime is related to the
number of distinct principle null directions (PND) of the Weyl tensor.
A generic spacetime will have four linearly independent null vectors
$k^\mu$ (i.e.\ PNDs) at all points that satisfy
\begin{equation}
 k^\nu k^\rho k_{[\tau}C_{\mu]\nu\rho[\sigma}k_{\chi]} = 0.
\label{eq:weylpnd}
\end{equation}
Type I spacetimes have four distinct PNDs, Type II have three distinct
PNDs (1 pair and two additional distinct PNDs), Type III have two
distinct PNDs with one PND of multiplicity three, Type D spacetimes
have two distinct PNDs consisting of two pairs of PNDs of multiplicity
two, type N spacetimes have a single PND of multiplicity four, and
Type O spacetimes have $C_{\mu \nu \rho \sigma} = 0$.

If the tetrad is chosen such that $l^a$ is a PND, then the Weyl scalar
$\psi_0 = C_{\mu \nu \rho \sigma} l^{\mu} m^{\nu} l^{\rho} m^{\sigma}$
vanishes, and similarly, if $\psi_0=0$, then $l^a$ is a PND.  Hence the
algebraic classification of the spacetime can be obtained by finding
the number of distinct choices of  $l^a$ for which $\psi_0=0$.  This
amounts to finding the roots (and multiplicity of the roots) of the
quartic equation (See Ref.~\cite{Stephani:2003tm}, Eq. (9.5))
\begin{equation}\label{eq:lambda}
\psi_0+4\lambda\psi_1+6\lambda^2\psi_2+4\lambda^3\psi_3+\lambda^4\psi_4=0,
\end{equation}
where $\psi_0,...,\psi_4$ are the Weyl scalars in an arbitrary tetrad,
restricted only by the condition $\psi_4\neq0$.
 This is equivalent to finding a tetrad
rotation such that $\psi_0=0$, and if the root is
repeated, then in this tetrad,  $\psi_1=0$ 
(similarly if the multiplicity of the root is 3 or 4 then
$\psi_2=0$ and $\psi_3=0$ respectively).
If, as in type D spacetimes, there are two pairs of
repeated PNDs, then we can choose a tetrad where the only non-vanishing
Weyl scalar is $\psi_2$.  It is important to note that the algebraic
classification is done pointwise. A spacetime, as a whole, is of a
particular type, if at every point the algebraic classification is of
that type.

In order to determine if the numerical spacetime is algebraically
special (within the numerical errors) we follow~\cite{dInverno71}
and~\cite{Stephani:2003tm}, Ch. 4. We start by defining the scalar
invariants~\cite{Carminati91}
\begin{equation}\label{eq:IJC}
I=\frac12\tilde C_{\alpha\beta\gamma\delta}
     \tilde C^{\alpha\beta\gamma\delta} ~~{\rm and}~~ 
J=-\frac16\tilde C_{\alpha\beta\gamma\delta}
    \tilde {C^{\gamma\delta}}_{\!\mu\nu}\tilde C^{\mu\nu\alpha\beta}.
\end{equation}
where
$\tilde
C_{\alpha\beta\gamma\delta}=\frac14(C_{\alpha\beta\gamma\delta}+\frac{i}{2}\epsilon_{\alpha\beta\mu\nu}{C^{\mu\nu}}_{\!\gamma\delta})$
(i.e.\ 1/2 the conjugate of the self-dual part of the Weyl tensor
$C_{\alpha\beta\gamma\delta}$).

If a spacetime has repeated principal null directions it is
algebraically special.  If this is the case, Eq.~(\ref{eq:lambda}) has
at least two repeated roots. In any case, Eq.~(\ref{eq:lambda}) can be
transformed into a depressed quartic (see Eq.~(\ref{eq:x}) below)
that, in turn, can be converted into a depressed nested cubic with
roots $y$, which satisfy the condition
\begin{equation}\label{eq:y}
y^3-Iy+2J=0.
\end{equation}
Algebraic specialty then implies
\begin{equation}\label{eq:D}
I^3=27J^2,
\end{equation}
i.e.\ $S=1$ in Eq.~\ref{eq:S}.
For Types $II$ and $D$ the invariants $I$ and $J$ are non-trivial,
while for Types $III$, $N$, and $O$ they vanish identically.

For practical applications, it is convenient to write the invariants
in terms of Weyl scalars in an arbitrary null tetrad
\begin{eqnarray}\label{eq:IJ}
I&=&3{\psi_2 }^2-4\psi_1\psi_3+\psi_4\psi_0, \\
J&=&-\psi_2^3+\psi_0\psi_4\psi_2 + 2\psi_1\psi_3\psi_2-\psi_4\psi_1^2
    -\psi_0\psi_3^2.
\end{eqnarray}

In order to completely determine the algebraic type we 
reduce Eq.~(\ref{eq:lambda}), by changing to the variable 
$x=\lambda\,\psi_4+\psi_3$~\cite{Gunnarsen95}, to the form
\begin{equation}\label{eq:x}
x^4+6\,L\,x^2+4\,K\,x+N=0,
\end{equation}
where
\begin{eqnarray}\label{eq:KLN}
K&=&\psi_1\psi_4^2-3\psi_4\psi_3\psi_2+2\psi_3^3,\\
L&=&\psi_2\psi_4-\psi_3^2,\\
N&=&\psi_4^2I-3L^2\nonumber\\
&=&{\psi_{{4}}}^{3}\psi_{{0}}-4\,{\psi_{{4}}}^{2}\psi_{{1}}\psi_{{3}}+6\,
\psi_{{4}}\psi_{{2}}{\psi_{{3}}}^{2}-3\,{\psi_{{3}}}^{4}
\end{eqnarray}
(note the typo in the definition of $N$ in
Refs.~\cite{Stephani:2003tm, dInverno71}).
For a type $II$ spacetime, $K\neq0$ and $N-9L^2\neq0$, while for
type $D$ and $III$ spacetimes, $K=0$ and $N-9L^2=0$ with $N\neq0$.
For a type $N$ spacetime, $K=0$ and $L=0$ (hence $N=0$).

Note that the above scalar objects are not invariant under arbitrary
tetrad rotations (See Ref.~\cite{Chandrasekhar83}, Chapter 1,  Eqs.~(342), 
[note typo there], (346) and (347)). 
Tetrad rotations are classified as Type I, II, and III, and have the
form:
\begin{eqnarray}
  \nonumber l^\mu &\to& l^\mu,\\
  \nonumber n^\mu &\to& n^\mu + \bar a m^\mu + a \bar m^\mu + a\bar a l^\mu,\\
  \nonumber m^\mu &\to& m^\mu+ a l^\mu,\\
  \bar m^\mu &\to& \bar m^\mu+ \bar a l^\mu,
\end{eqnarray}  
\begin{eqnarray}
  \nonumber l^\mu &\to& l^\mu + \bar b m^\mu + b \bar m^\mu + b\bar b n^\mu,\\
  \nonumber n^\mu &\to& n^\mu,\\
  \nonumber m^\mu &\to& m^\mu+ b n^\mu,\\
  \bar m^\mu &\to& \bar m^\mu+ \bar b n^\mu,
\end{eqnarray}
\begin{eqnarray}
  \nonumber l^\mu &\to& A^{-1} l^\mu,\\
  \nonumber n^\mu &\to& A n^\mu,\\
  \nonumber m^\mu &\to& e^{i\theta} m^\mu,\\
  \bar m^\mu &\to& e^{-i\theta} \bar m^\mu,
\end{eqnarray}
for Type I, II, and III, respectively,
where $a$ and $b$ are complex scalars and $A$ and
$\theta$ are real scalars.
Under these rotations the scalars $L$, $K$, and $N$ transform as
\begin{eqnarray}
L&\to&A^2e^{-2I\theta}L,\nonumber\\
K&\to&A^3e^{-3I\theta}K,\nonumber\\
N&\to&A^4e^{-4I\theta}N.
\end{eqnarray}
for Type III rotations and
\begin{eqnarray}
L&\to&L,\nonumber\\
K&\to&K,\nonumber\\
N&\to&N,
\end{eqnarray}
for Type II rotations.
Expressions for Type I rotations do not have these simple
forms, but we verified
that, if as in type D solutions, $K=0$ and $N-9L^2=0$ in the original
tetrad, then $K=0$ and $N-9L^2=0$ in the new rotated tetrad
(this is also obvious for type III and II transformations above).
One the other hand $L=0$ is not preserved by
type I rotations.

Coming back to the roots $x_1,x_2,x_3,x_4$ of Eq.~(\ref{eq:x}), we
observe that, in numerically generated spacetimes, the roots  never
agree exactly, even if the metric is expected to be of a special
algebraic type. Of course, the root differences in each pair should
scale with resolution and asymptotically approach zero as $h\to 0$ and
$t\to\infty$ (where $h$ is the gridspacing).

The roots of Eq.~(\ref{eq:lambda}) can be obtained from the roots  of
Eq.~(\ref{eq:y})
using the following algorithm~\cite{AbramowitzStegun}
\begin{eqnarray}\label{eq:yroots-alt}
D&=&J^2-(I/3)^3,\nonumber\\
A&=&(-J+\sqrt{D})^{1/3},\quad B=(-J-\sqrt{D})^{1/3}\nonumber,\\
y_1&=&A+B,\nonumber\\
y_2&=&-\frac12(A+B)+i\frac{\sqrt{3}}{2}(A-B),\nonumber\\
y_3&=&-\frac12(A+B)-i\frac{\sqrt{3}}{2}(A-B),
\end{eqnarray}
where  the complex phases of $A$ and $B$ are chosen such that  $A\,B=I/3$.
The roots of Eq.\ (\ref{eq:x}) are then obtained from the roots of 
the complete cubic equation for the
variable $z$ (where $z=2\psi_4\,y-4L$)
\begin{equation}
z^3+12\,L\,z^2+4(9L^2-N)\,z-16K=0,
\end{equation}
which has the roots
\begin{eqnarray}\label{eq:zroots}
z_1&=&2\psi_4\,y_1-4L,\nonumber\\
z_2&=&2\psi_4\,y_2-4L,\nonumber\\
z_3&=&2\psi_4\,y_3-4L.
\end{eqnarray}
Finally the roots of our original equation (\ref{eq:lambda}) can
be written in the form~\cite{Gunnarsen95}
\begin{eqnarray}\label{eq:lambdaroots}
\lambda_1&=&\left[-\psi_3+\frac12(\sqrt{z_1}+\sqrt{z_2}+\sqrt{z_3})\right]/\psi_4,\nonumber\\
\lambda_2&=&\left[-\psi_3+\frac12(\sqrt{z_1}-\sqrt{z_2}-\sqrt{z_3})\right]/\psi_4,\nonumber\\
\lambda_3&=&\left[-\psi_3+\frac12(-\sqrt{z_1}+\sqrt{z_2}-\sqrt{z_3})\right]/\psi_4,\nonumber\\
\lambda_4&=&\left[-\psi_3+\frac12(-\sqrt{z_1}-\sqrt{z_2}+\sqrt{z_3})\right]/\psi_4,
\end{eqnarray}
where the signs of the $\sqrt{z_i}$ are chosen such that 
$(\sqrt{z_1}\sqrt{z_2}\sqrt{z_3})=-4K$.
We note that in a type D spacetime 
$\lambda_1=\lambda_2$ and $\lambda_3=\lambda_4$.

\subsection{Vacuum}

The determination of the algebraic type of the matter fields can be
done in an analogous way using the Ricci tensor, rather than the
Weyl scalars.
The analogue of the Petrov
types are the Segre types and the equation to determine the multiplicities
of the roots is~(\cite{Stephani:2003tm}, Eq. (9.2))
\begin{equation}\label{eq:sigma}
\sigma^4-\frac12I_6\,\sigma^2-\frac13I_7\,\sigma+\frac18(I_6^2-2I_8)=0,
\end{equation}
where
\begin{eqnarray}\label{eq:I6I7I8}
I_6&=&S^\alpha_\beta\,S^\beta_\alpha,\\
I_7&=&S^\alpha_\beta\,S^\beta_\gamma\,S^\gamma_\alpha,\\
I_8&=&S^\alpha_\beta\,S^\beta_\gamma\,S^\gamma_\delta\,S^\delta_\alpha,
\end{eqnarray}
and
\begin{equation}\label{eq:Sab}
S_{\alpha\beta}=R_{\alpha\beta}-\frac14g_{\alpha\beta}R,
\end{equation}
is the trace free part of the Ricci tensor.

This characterization of the matter fields does not completely
determine the algebraic properties, and other additional criteria have
to be used. In our numerical simulations here, we are concerned with
vacuum spacetimes. Numerical evolutions may introduce artificial (and
unphysical) matter fields through violations of the Hamiltonian and
momentum constraints, and  the natural way of monitoring the accuracy
of the solution is to examine these constraints and confirm that the
induced matter fields converge to zero.

\subsection{Determination of the Kerr solution}

Once we determine that a solution is, for instance, Petrov type D and is
a vacuum solution, we still do not uniquely single out the Kerr spacetime.
One can go further and try to determine if the spacetime has the
symmetries of Kerr (the Kerr spacetime
has two commuting spacelike and timelike Killing vectors
\cite{Stephani:2004ac}). However, one still needs to examine the 
 asymptotic behavior of the solutions to determine that
the spacetime does not have a NUT charge $l$ or acceleration $\alpha$. 

A general type D, vacuum Black hole solution can be described by the metric
(\cite{Griffiths:2005se}, Eq. (17)),

\begin{eqnarray}\label{eq:metric}
&&ds^2=\frac{1}{\Omega^2}\left\{
\frac{Q}{\rho^2}\left[dt- \left(a\sin^2\theta
+4l\sin^2{\textstyle{\theta\over2}} \right)\,d\phi \right]^2
   -{\rho^2\over Q}\,dr^2 \right.\nonumber\\
&& \left. -{P\over\rho^2} \Big[a dt  
  -\Big(r^2+(a+l)^2\Big)d\phi \Big]^2  
-{\rho^2\over P}\sin^2\theta\,d\theta^2 \right\}, 
\end{eqnarray}
where 
\begin{eqnarray}
\Omega&=&1-{\alpha}(l+a\cos\theta)\,r,\\
\rho^2&=&r^2+(l+a\cos\theta)^2,\\
P&=& \sin^2\theta\,(1-a_3\cos\theta-a_4\cos^2\theta),\\
Q&=&k -2mr +\epsilon r^2-2\alpha{n} r^3 -\alpha^2kr^4,
\end{eqnarray}
and 
\begin{eqnarray}
a_3&=& 2\alpha{a}m -4\alpha^2\,a\,l\,k,\\
a_4&=& -\alpha^2\,a^2\,k
\end{eqnarray}
with $\epsilon$, $n$ and $k$ as given a function of the more basic parameters
$m$, $l$, $a$, and $\alpha$ by 
\begin{eqnarray}
 &&\epsilon= {k\over a^2-l^2}+4\alpha lm 
 -(a^2+3l^2) {\alpha^2}k, \label{epsilon}\\
 &&n= {k\,l\over a^2-l^2} -\alpha{(a^2-l^2)}\,m 
 +(a^2-l^2)l \alpha^2k,
  \label{n}  \\
&&  \left( {1\over a^2-l^2}+3\alpha^2l^2 \right)\,k 
  =1 +2\alpha lm .
  \label{k}
\end{eqnarray}

If the null tetrad is aligned with the principal null
directions, i.e.\

 \begin{equation}
 \begin{array}{l}
 l^\mu ={\displaystyle {(1-\alpha pr)\over\sqrt{2(r^2+p^2)}}\left[ 
 {1\over\sqrt Q}\Big(r^2\partial_\tau-\partial_\sigma\Big) 
-\sqrt Q\,\partial_r \right]} \,, \\[16pt] 
 n^\mu ={\displaystyle {(1-\alpha pr)\over\sqrt{2(r^2+p^2)}}\left[ 
 {1\over\sqrt Q}\Big(r^2\partial_\tau-\partial_\sigma\Big) 
+\sqrt Q\,\partial_r \right]} \,, \\[16pt] 
 m^\mu ={\displaystyle {(1-\alpha pr)\over\sqrt{2(r^2+p^2)}}\left[ 
 -{1\over\sqrt P}\Big(p^2\partial_\tau+\partial_\sigma\Big) 
+i\sqrt P\,\partial_p \right]} \,,  
 \end{array} 
 \label{GeneralTetrad}
 \end{equation}

then the only non-vanishing Weyl scalar is
\begin{equation}\label{eq:psi2}
 \Psi_2=-(m+in)\left({1-\alpha pr\over r+ip}\right)^3.
\end{equation}
where $p=l+a\cos\theta$.

It is then natural to look at the asymptotic behavior of the spacetime
to determine if there is a NUT charge $l$, an acceleration $\alpha$, 
or if the spacetime is plain
Kerr. One can use the method of determining a quasi-Kinnersley
frame~\cite{Beetle:2004wu, Campanelli:2005ia} to compute $\psi_2$ and
perform the above analysis.  Alternatively, we can use the fact that,
once we determined the spacetime is type D, we can choose a tetrad
where all the Weyl scalars, but $\psi_2$, vanish. Hence the invariants $I$
and $J$ must have the form
\begin{equation}\label{eq:IJpnd}
I=3\psi_2^2,\quad J=-\psi_2^3
\end{equation}
in this special class of tetrads.

If the acceleration $\alpha\not=0$ then a series expansion of the
invariant $I$ gives
\begin{equation}\label{eq:IexpansionGeneral}
I=3(m+il)^2\alpha^6p^6-\frac{18}{r}(m+il)^2\alpha^5p^5
(i\alpha p^2+1)+{\cal O}
\left(\frac{1}{r^2}\right).
\end{equation}

Note that if the acceleration $\alpha=0$ then $n=l$.
An asymptotic expansion of the $I$ invariant for the metric~(\ref{eq:metric})
then gives
\begin{equation}\label{eq:Iexpansion}
I=\frac{3}{r^6}(m+il)^2-\frac{18i}{r^7}(m+il)^2(l+a\cos\theta)+{\cal O}
\left(\frac{1}{r^8}\right),
\end{equation}
and, by looking at the real and imaginary parts of the $I$ invariant at
large radii, we can determine the $l$ parameter via
\begin{eqnarray}\label{eq:ReIm}
\Im(I)/\Re(I)=
\begin{cases}
\frac{2ml}{m^2-l^2};\quad l\not=0,\\
\frac{-6a\cos\theta}{r};\quad l=0.\\
\end{cases}
\end{eqnarray}
We will use this method to determine the asymptotic behavior
of the final remnant of a black-hole-binary merger.
Note that using $I$ and $J$ only requires smooth second derivatives of
the metric, which has a distinct advantage over higher-derivative
methods when dealing with numerically
generated spacetimes.

\section{Numerical Techniques}
\label{sec:nrtechniques}
To compute the numerical initial data, we use the puncture
approach~\cite{Brandt97b} along with the {\sc
TwoPunctures}~\cite{Ansorg:2004ds} code.  In this approach the
3-metric on the initial slice has the form $\gamma_{a b} = (\psi_{BL}
+ u)^4 \delta_{a b}$, where $\psi_{BL}$ is the Brill-Lindquist
conformal factor, $\delta_{ab}$ is the Euclidean metric, and $u$ is
(at least) $C^2$ on the punctures.  The Brill-Lindquist conformal
factor is given by $ \psi_{BL} = 1 + \sum_{i=1}^n m_{i}^p / (2 |\vec r
- \vec r_i|), $ where $n$ is the total number of `punctures',
  $m_{i}^p$ is the mass parameter of puncture $i$ ($m_{i}^p$ is {\em
not} the horizon mass associated with puncture $i$), and $\vec r_i$ is
the coordinate location of puncture $i$.  We evolve these
black-hole-binary data-sets using the {\sc
LazEv}~\cite{Zlochower:2005bj} implementation of the moving puncture
approach~\cite{Campanelli:2005dd, Baker:2005vv}.  In our version of
the moving puncture approach we replace the
BSSN~\cite{Nakamura87,Shibata95, Baumgarte99} conformal exponent
$\phi$, which has logarithmic singularities at the punctures, with the
initially $C^4$ field $\chi = \exp(-4\phi)$.  This new variable, along
with the other BSSN variables, will remain finite provided that one
uses a suitable choice for the gauge. An alternative approach uses
standard finite differencing of $\phi$~\cite{Baker:2005vv}.  Recently
Marronetti {\it et al.}~\cite{Marronetti:2007wz} proposed the use of
$W=\sqrt{\chi}$ as an evolution variable.  For the runs presented here
we use centered, eighth-order finite differencing in
space~\cite{Lousto:2007rj} and a fourth-order Runge-Kutta time integrator (note that we do
not upwind the advection terms).

We use the Carpet~\cite{Schnetter-etal-03b} mesh refinement driver to
provide a `moving boxes' style mesh refinement. In this approach
refined grids of fixed size are arranged about the coordinate centers
of both holes.  The Carpet code then moves these fine grids about the
computational domain by following the trajectories of the two black
holes.

We obtain accurate, convergent waveforms and horizon parameters by
evolving this system in conjunction with a modified 1+log lapse and a
modified Gamma-driver shift condition~\cite{Alcubierre02a,
Campanelli:2005dd}, and an initial lapse $\alpha(t=0) =
2/(1+\psi_{BL}^{4})$.  The lapse and shift are evolved with
\begin{subequations}
\label{eq:gauge}
  \begin{eqnarray}
(\partial_t - \beta^i \partial_i) \alpha &=& - 2 \alpha K,\\
 \partial_t \beta^a &=& B^a, \\
 \partial_t B^a &=& 3/4 \partial_t \tilde \Gamma^a - \eta B^a.
 \label{eq:Bdot}
 \end{eqnarray}
\end{subequations}
These gauge conditions require careful treatment of $\chi$, the
inverse of the three-metric conformal factor, near the puncture in
order for the system to remain stable~\cite{Campanelli:2005dd,
Campanelli:2006gf, Bruegmann:2006at}.  As shown in
Ref.~\cite{Gundlach:2006tw}, this choice of gauge leads to a strongly
hyperbolic evolution system provided that the shift does not become
too large.  In our tests, $W$ showed better behavior at very early
times ($t < 10M$) (i.e.\ did not require any special treatment near
the punctures), but led to evolutions with larger truncation errors
(importantly, larger orbital phase errors) 
when compared to $\chi$.

We use {\sc AHFinderDirect}~\cite{Thornburg2003:AH-finding} to locate
apparent horizons.  We measure the magnitude of the horizon spin using
the Isolated Horizon algorithm detailed in~\cite{Dreyer02a}. This
algorithm is based on finding an approximate rotational Killing vector
(i.e.\ an approximate rotational symmetry) on the horizon $\varphi^a$.
Given this approximate Killing vector $\varphi^a$, the spin magnitude
is
\begin{equation}
 \label{eq:isospin} S_{[\varphi]} =
 \frac{1}{8\pi}\oint_{AH}(\varphi^aR^bK_{ab})d^2V,
\end{equation}
where $K_{ab}$ is the extrinsic curvature of the 3D-slice, $d^2V$ is
the natural volume element intrinsic to the horizon, and $R^a$ is the
outward pointing unit vector normal to the horizon on the 3D-slice.
We measure the direction of the spin by finding the coordinate line
joining the poles of this Killing vector field using the technique
introduced in~\cite{Campanelli:2006fy}.  Our algorithm for finding the
poles of the Killing vector field has an accuracy of $\sim 2^\circ$
(see~\cite{Campanelli:2006fy} for details). Note that once we have the
horizon spin, we can calculate the horizon mass via the Christodoulou
formula (which is exact for a Kerr black-hole)
\begin{equation}
 \label{eq:isomass}
{m^H} = \sqrt{m_{\rm irr}^2 +
 S^2/(4 m_{\rm irr}^2)},
\end{equation}
where $m_{\rm irr} = \sqrt{A/(16 \pi)}$ and $A$ is the surface area of
the horizon.

We also use an alternative quasi-local measurement of the spin and
linear momentum of the individual black holes in the binary that is
based on the coordinate rotation and translation
vectors~\cite{Krishnan:2007pu}.  In this approach the spin components
of the horizon are given by
\begin{equation} S_{[i]} =
\frac{1}{8\pi}\oint_{AH} \phi^a_{[i]} R^b K_{ab} d^2V,
\label{eq:coordspin}
\end{equation}
 where $\phi^i_{[\ell]} =
\delta_{\ell j} \delta_{m k} r^m \epsilon^{i j k}$, $\epsilon^{1\ 2\ 3}=1$,
 and $r^m = x^m -
x_0^m$ is the coordinate displacement from the centroid of the hole,
while the linear momentum is given by
\begin{equation} P_{[i]} =
\frac{1}{8\pi}\oint_{AH} \xi^a_{[i]} R^b (K_{ab} - K \gamma_{ab})
d^2V, \label{eq:coordmom}
 \end{equation}
 where $\xi^i_{[\ell]} = \delta^i_\ell$.

\subsection{Numerical Tetrad and Root Finder}
\label{sec:tetrad}
We calculate $\psi_0\cdots\psi_4$ using the tetrad 
\begin{eqnarray}
l^\mu &=& (t^\mu + r^\mu)/\sqrt{2},\\
n^\mu &=& (t^\mu - r^\mu)/\sqrt{2},\\
m^\mu &=& (\theta^\mu + i \phi^\mu)/\sqrt{2},
\end{eqnarray}
where $t^\mu$ is the unit normal to the $t={\rm const}$ slices and
$\{r^\mu,\theta^\mu,\phi^mu\}$ are unit spacelike vectors (with time
component equal to zero) constructed as follows~\cite{Baker:2001sf}.
We start with the unit vector
\begin{equation}
  \phi^a = \widehat {\tilde \phi^a},
\end{equation}
where $\tilde \phi^a = \{-y, x, 0\}$ ,
$\widehat{v^a} = v^a /\sqrt{v^a v^b \gamma_{ab}}$, and
$\gamma_{ab}$ is the spatial metric. We then find the
unit vector in radial direction perpendicular to $\phi^a$
\begin{equation}
 r^a = \widehat{\tilde r^a},
\end{equation}
where
\begin{equation}
 \tilde r^a = \breve r^a - \breve r^a \phi^b \gamma_{ab},
\end{equation}
and $\breve r^a = \{x,y,z\}$. Finally, we obtain
\begin{equation}
  \theta^a = \widehat{\tilde \theta^a},
\end{equation}
where
\begin{equation}
  \tilde \theta^a = \gamma^{a b} \epsilon_{b c d} \phi^c r^d.
\end{equation}
With this choice of tetrad $\psi_0\cdots\psi_4$ are all non-vanishing for
Kerr spacetimes when the specific spin $a$ is non-vanishing.

\subsection{Initial Data}
\label{sec:ID}

To generate the initial data parameters, we used random values for the
mass ratio and spins of the binary (the ranges for these parameters
were chosen to make the evolution practical). We then calculated
approximate quasi-circular orbital parameters for a binary with these
chosen parameters at an initial orbital separation of $50M$ and
evolved using purely PN evolutions until the binary separation
decreased to $2.3M$.  The goal was to produce a binary that had no
particular symmetries, so that we can draw general conclusions from
the results, while also merging very quickly (within $15M$ of the
start of the simulation), to reduce the computational expense.  The
initial binary configuration at $r=50M$ was chosen such that
$q=m_1/m_2 = 0.8$, $\vec S_1/m_1^2 = (-0.2, -0.14, 0.32)$, and $\vec
S_2/m_2^2 =(-0.09, 0.48, 0.35)$. This is the same basic configuration
that we used in~\cite{Campanelli:2008nk}.  We summarize the initial
data parameters in Table~\ref{table:ID}.
\begin{table}
\caption{Initial data parameters for the numerical evolution.
 The punctures have
mass parameters $m^p_i$, horizons masses $m^H_i$,
momenta $\pm\vec p$, spins $\vec S_i$, and the configuration has a
total ADM mass $M_{\rm ADM} = 1.0000004 M$. }
\label{table:ID}
\begin{ruledtabular}
\begin{tabular}{lccc}
$m^p_1/M$     & 0.37752 & $m^p_2/M$     & 0.42452\\
$m^H_1/M$     & 0.46298 & $m^H_2/M$     & 0.57872\\
$x_1/M$     &  -0.75023 & $x_2/M$       & 0.58004\\
$y_1/M$     &   1.11679 & $y_2/M$       &-0.89449\\
$z_1/M$     &  -0.16093 & $z_2/M$       & 0.20338\\
$S^x_1/M^2$ &  -0.020765& $S^x_2/M^2$   & 0.12106\\
$S^y_1/M^2$ &   0.065806& $S^y_2/M^2$   &-0.05532\\
$S^z_1/M^2$ &   0.054697& $S^z_2/M^2$   & 0.16178\\
$p^x/M$     &  -0.134735& $p^y/M$      & -0.21376\\
$p^z/M$     &  -0.012323 \\
\end{tabular}
\end{ruledtabular}
\end{table}

\section{Results}
\label{sec:results}

We ran the binary configuration using 9 levels of refinement with 
an outer grid of resolution $h=3.2M$ extending to $\pm416M$. The
resolution on the finest grid was $h=M/80$. We analyze the Weyl
scalars in the region $r \lesssim 5M$ where we had a resolution of $h\leq
M/20$. This calculation is non-trivial because the magnitudes of the
Weyl scalars can be quite small (we need to analyze these scalars at
very late times when the waveform amplitudes are quite small),
 requiring very-high overall
simulation accuracy. We found that the isolated horizon formulae and
the radiated energy and angular momentum both predict similar remnant
masses and spins, with the isolated horizon formulae
Eqs.~(\ref{eq:isospin})-(\ref{eq:coordspin}) giving
$M_{\rm rem} = 0.9859$, $\vec S_{\rm rem}=\{0.00160\pm0.00005,
0.0407\pm0.0004, 0.7173\pm0.0001\}$ and the radiation giving
$M_{\rm rem}= 0.9861\pm0.0001$, $\vec S_{\rm rem} = \{0.00153\pm0.00001,
0.04078\pm0.00002, 0.7179\pm0.0001\}$. A fit to the quasi-normal
profile $\sim \exp(-\alpha t) \sin(\omega t)$ gives
$\alpha = 0.07997\pm0.0013$ and $\omega = 0.5603\pm0.0025$, where the 
values quoted are the average from fits to the real and imaginary
parts of the $(\ell=2,m=2)$ component of $\psi_4$ extracted at
$r=100M$ over the domain $(160M < t < 200M)$. The resulting values of 
$M_{\rm rem}$ and $a/M_{\rm rem}$~\cite{Echeverria89} are
$0.9876\pm0.0079$ and $0.743\pm0.013$ respectively. Note that the
isolated horizon and radiated Energy/Momentum formulae predict that
the final specific spin is $a/M_{\rm rem} = 0.73931\pm0.00016$. This
agreement is consistent with the final remnant being a Kerr hole (Note
that this consistency is not a proof that the remnant is
Kerr).

\begin{figure}
\includegraphics[width=3.5in]{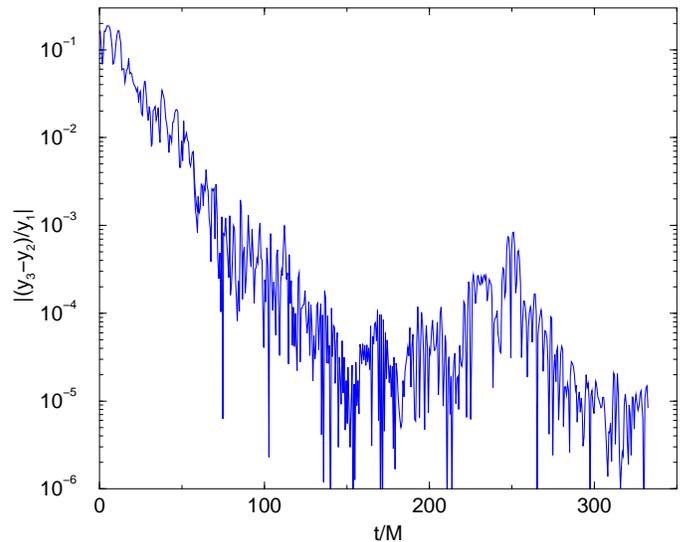}
\caption{The magnitude  $|(y_3-y_2)/y_1|$ versus time
at the point
$x=5M, y=0, z=0$. The spacetime is algebraically special if
$|(y_3-y_2)/y_1| = 0$. Note the initial exponential decrease in the
root difference.}
\label{fig:yratio}
\end{figure}

\begin{figure}
\includegraphics[width=3.5in]{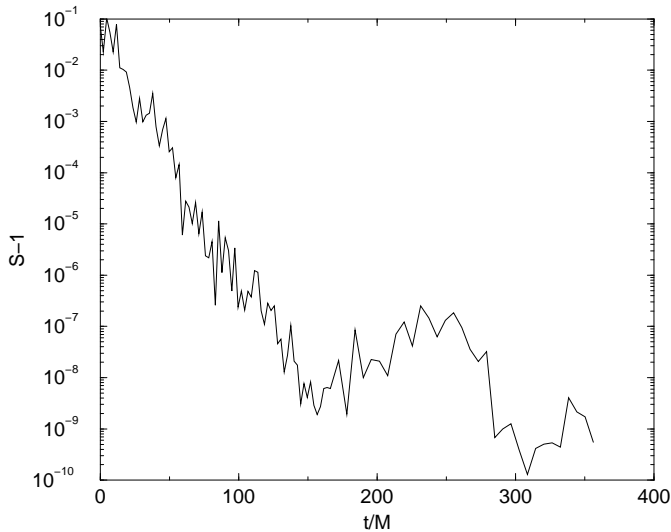}
\caption{The magnitude  $|S-1|$ versus time
at the point
$x=5M, y=0, z=0$. The spacetime is algebraically special if
$S = 1$. Note the initial exponential decrease in $S-1$.}
\label{fig:sinvar}
\end{figure}

If the spacetime is algebraically special, then the roots $y_2$ and
$y_3$ of Eq.~(\ref{eq:y}) are equal. To measure how far the spacetime
is from being algebraically special we plot the magnitude
$|(y_3-y_2)/y_1|$ (here $y_1$ provides a natural normalization) 
and the invariant $S-1$ at the point $(x=5M, y=z=0)$
(See Figs.~\ref{fig:yratio}~and~\ref{fig:sinvar})~\cite{Baker:2001sf,
Baker:2002qf, Campanelli:2005ia}.
From the figures we can see that the deviation of the spacetime from
being algebraically special decreases exponentially (with an e-folding
time of $\sim 20M$ for $y_1 - y_2$ and $\sim 10M$ for $S-1$)
 with time until $t\sim 150M$.
The oscillation seen after this time may be due to reflections off of
the refinement boundaries (this, in turn, provides a sensitive test to
improve the numerical techniques).

\begin{figure}
\includegraphics[width=3.5in]{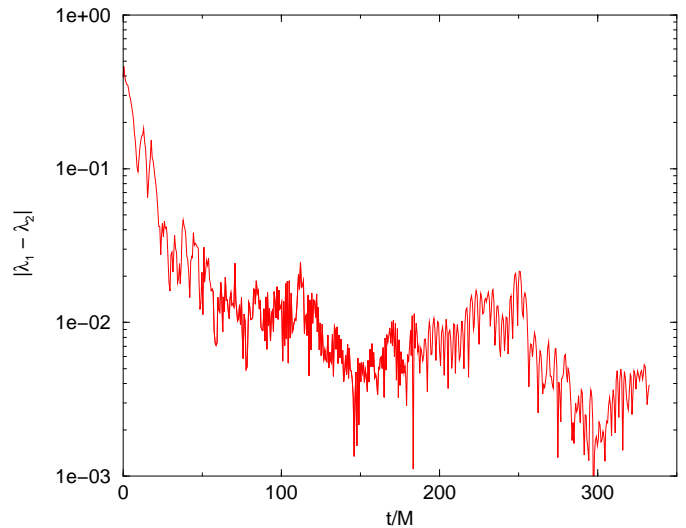}
\caption{The magnitude of the root-pair separation
$|\lambda_1 - \lambda_2|$ versus time
for the two roots close to $\lambda = 0$ at the point
$x=5M, y=0, z=0$.}
\label{fig:close_v_t}
\end{figure}

\begin{figure}
\includegraphics[width=3.5in]{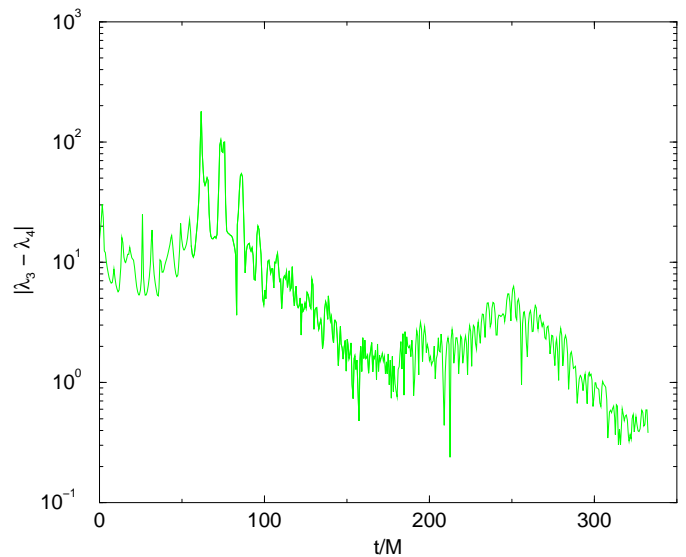}
\caption{The magnitude of the root-pair separation
$|\lambda_3 - \lambda_4|$ versus time
for the two roots furthest from $\lambda = 0$ at the point
$x=5M, y=0, z=0$. Note that there is no rapid decrease in the
$|\lambda_3 - \lambda_4|$ which indicates that the spacetime is not
approaching type D as fast as it is approaching type II.} 
\label{fig:far_v_t}
\end{figure}

\begin{figure}
\includegraphics[width=3.5in]{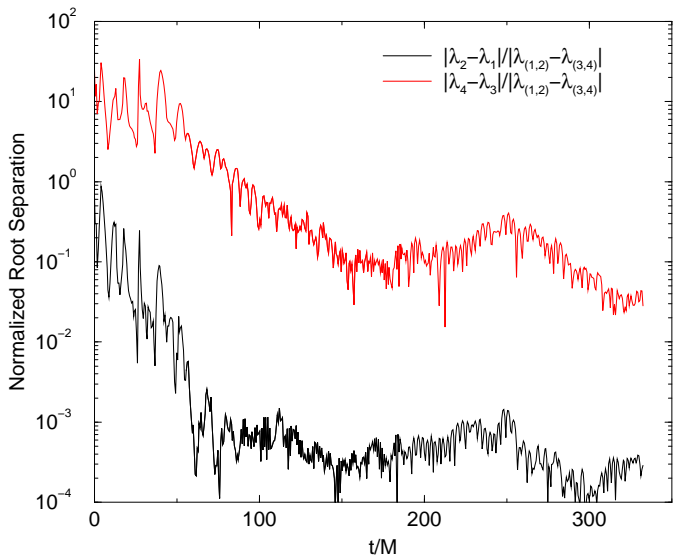}
\caption{The magnitude of the two root-pair separations normalized by
the magnitude of the differences of the average value of the roots in
each pair $|\lambda_{(1,2)} - \lambda_{(3,4)}|$, where
 $\lambda_{(1,2)} = (\lambda_1+\lambda_2)/2$  and
 $\lambda_{(3,4)} = (\lambda_3+\lambda_4)/2$.
}
\label{fig:normalized_root_sep}
\end{figure}

\begin{figure}
\includegraphics[width=3.5in]{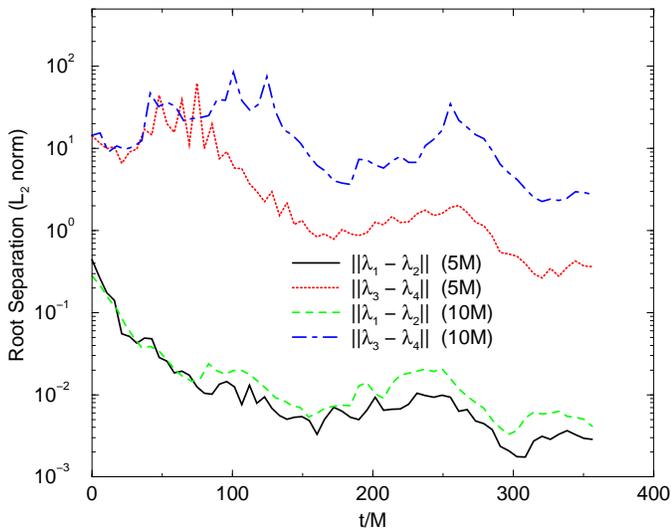}
\caption{The $L_2$ norm of the root separations versus time along the
$x$ and $y$ axis for $2<|x|,|y|<5$ and $2<|x|,|y|<10$. The region
containing the black hole itself was excluded from the norm.}
\label{fig:root_norm}
\end{figure}

\begin{figure}
\includegraphics[width=3.5in]{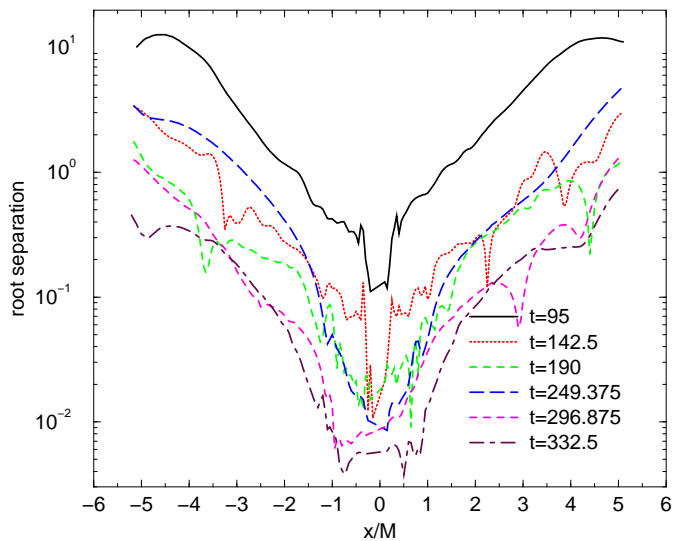}
\caption{The magnitude of the root-pair separation
$|\lambda_3 - \lambda_4|$ along the $x$-axis for
several values of $t$. At first the root separation decreases
significantly with time, but eventually stabilizes as numerical errors
due to reflections off the refinement boundaries and other numerical
sources of error begin to dominate.
}
\label{fig:far_v_x_t_all}
\end{figure}

\begin{figure}
\includegraphics[width=3.5in]{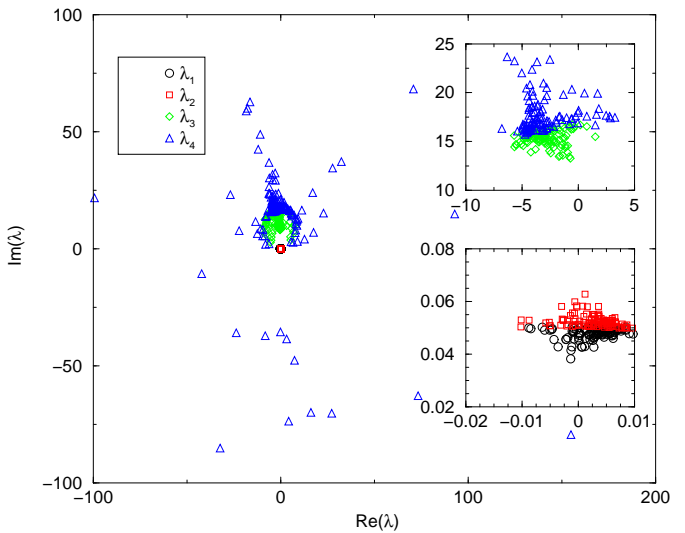}
\caption{The locations on the complex plane of the roots
$\lambda_1,\cdots, \lambda_4$
for $t=57, 57.59375, \cdots,
166.25$ at the point $(x=5, y=0, z=0)$.
 The insets shows the last 107 points. Note that $\lambda_4$
has the largest scatter in time and that the separation of $\lambda_1$
and $\lambda_2$ is not distinguishable on the overall plot. Initially,
the points at different times are scattered, but converge to a fixed
limit at late times.
}
\label{fig:root_plane}
\end{figure}

In Figs.~\ref{fig:close_v_t}-\ref{fig:far_v_x_t_all} we show the
unnormalized magnitudes of the root-pair differences $|\lambda_1 - \lambda_2|$ and
$|\lambda_3 - \lambda_4|$ both as a function of $t$ at a fixed
$(x,y,z) = (5,0,0)$  and along the $x$-axis at several times.
Both pairs show a general decrease in the magnitudes of the
differences with time, but with a pronounced oscillatory behavior.
Note that $|\lambda_1 - \lambda_2|$ separation is much smaller than
the $|\lambda_3 - \lambda_4|$ separation, indicating that the
space-time first approaches Type II (and hence is algebraically
special with $S-1\sim 0$) before settling to Type D.  In
Fig.~\ref{fig:root_plane} we plot
the values of the pairs $(\lambda_1, \lambda_2)$ and $(\lambda_3,
\lambda_4)$ on the complex plane at the point $(5,0,0)$ for times
$t=57,\cdots, 166.25$ in steps of $0.59375$. From the plots we can see how each
of the two roots in the root pairs approach each other.
In Fig.~\ref{fig:normalized_root_sep} we plot the magnitude of the root
separations normalized by the difference between the average value of
the roots in each pair (note that $|\lambda_2 - \lambda_3|$ has an
e-folding time of $\sim30M$). It takes about $80M$ of evolution, or $65 M$
post merger, until the larger normalized root separation falls below 1.
Finally, in Fig.~\ref{fig:root_norm} we show the $L_2$ norm of the root
separations along the $x$ and $y$ axes restricted to $2M<|x|,|y|<5M$
and $2M < |x|,|y| < 10M$ (the restriction to $|x|,|y|>2M$ is such that
the black-hole interior is not included in the norm). The poorer
convergence of the norm over the larger domain is due to numerical
errors in the more coarsely resolved regions. 


\begin{figure}
\includegraphics[width=3.5in]{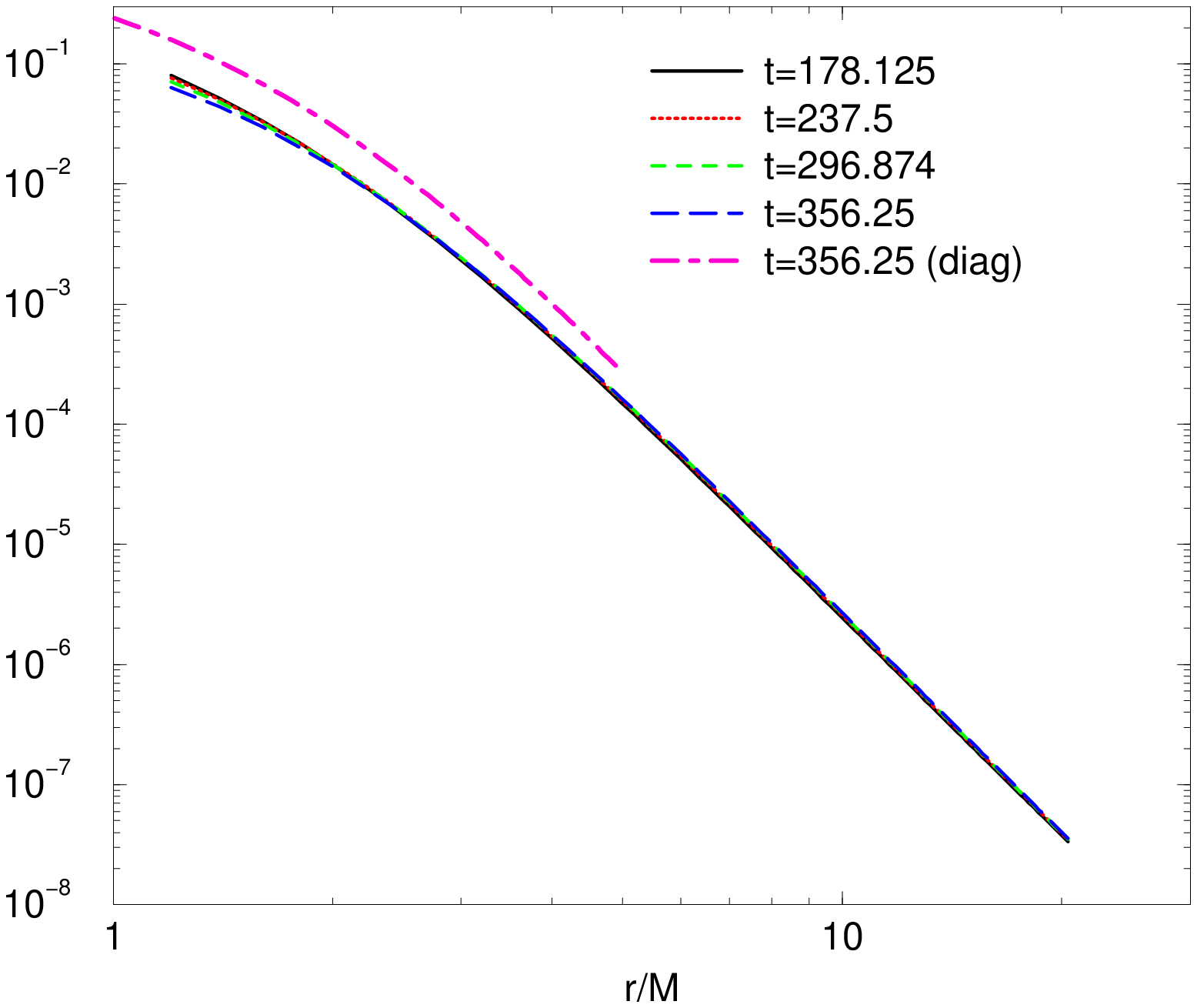}
\caption{$r |I|$ as a function of
$r/M$ along the $y$-axis and the diagonal line $(x=0,y=z)$.
 Note that the behavior indicates that
$ r|I| \to 0$ as $r\to \infty$,
which indicates that the acceleration $\alpha$ 
vanishes.}
\label{fig:rI_v_y}
\end{figure}

\begin{figure}
\includegraphics[width=3.5in]{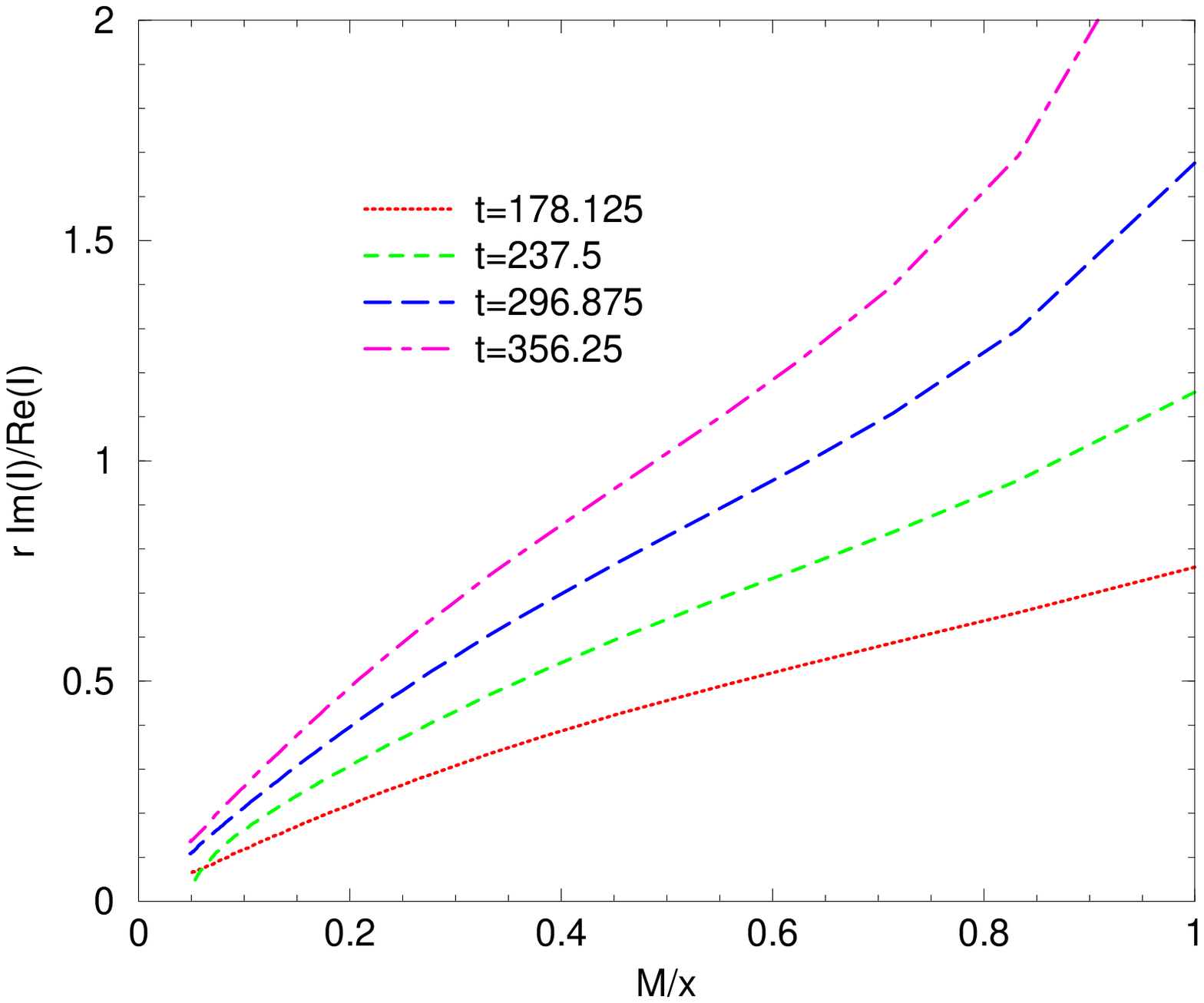}
\caption{The ratio $r \Im(I)/\Re(I)$ as a function of
$M/x$ along the $x$-axis. Note that the behavior indicates that
$\Im(I)/\Re(I)\to 0$ as $r\to\infty$ (i.e.\ $M/x\to 0$),
which indicates that the NUT
charge vanishes.}
\label{fig:invarratio}
\end{figure}

In Fig.~\ref{fig:rI_v_y} we plot $r |I|$ versus $r/M$ along the
$+y$-axis and along the line $(x=0, y=z)$.  The leading-order term if
$\alpha\neq0$ and $l=0$  has an $(\alpha p)^6$ dependence, where $p=l+a
\cos\theta$ (See
Eq.~(\ref{eq:metric})). If $l=0$ then $p=a \cos \theta$, and along the
$y$-axis, $p^6\sim 10^{-9}$ (the remnant spin is slightly misaligned
with the $z$-axis), but along the line $(x=0, y=z)$,
$p^6\sim0.028$. From the data on the $y$-axis we can only conclude
that $\alpha l$ is very small. However, along the diagonal,
$p^6 \sim .028 + 0.30 l$, which provides evidence that both
 $\alpha$ and $l$
are small.
In Fig.~\ref{fig:invarratio} we plot the function $r \Im(I)/\Re(I)$
versus $M/x$ along the $+x$-axis for various times from 
$t\sim 100$ to $t\sim350M$.
It is clear from the plot that this function does not tend to $\infty$
at larger $r$, which indicates that the NUT charge of the space time
vanishes (i.e.\ given that we already found that $\alpha$ vanishes).
Hence we can see good evidence that the spacetime is approaching Type
D with zero NUT charge and zero acceleration, and hence is approaching 
a Kerr spacetime.

 We have
confirmed that the constraints converge to zero for our code outside
of the horizons. For this simulation the constraint violations where
of order $10^{-4}$ at the horizons, and dropped off steeply with
radius. Convergence of the constraints is important to show that the
spacetime remains a vacuum spacetime outside of the remnant horizons.

\section{Conclusion}
\label{sec:discussion}

We have provided a method to classify numerically generated
spacetimes according to their algebraic properties. This is 
based on the use of the coincidence of the principal null directions
for algebraically special spacetimes. In particular, we focus on 
the final remnant of a generic-black-hole-binary merger,
that, according to the `no hair' theorem, is expected to produce a
Kerr black hole, and hence be of algebraic (Petrov) type D 
(i.e.\ that the four principal null directions agree in pairs). We give
a measure of the agreement by normalizing the numerical differences
between two nearby roots of Eq.~(\ref{eq:lambda}) with the average separation
to the other root pair in the complex plane.

We have been able to verify this agreement to order $10^{-4}$
and $10^{-2}$ for the two pairs respectively. We find that the 
agreement of the two roots in each pair improves with evolution time
and only appears to be limited by unphysical boundary effects
(from the refinement and outer boundaries). The late-time behavior
of these two root pairs implies that the spacetime near the remnant
first approaches an algebraically special type II (with one pair of roots 
and two
distinct roots) and over longer timescales approaches type D. We also
analyze the invariant asymptotic behavior of the spacetime and do
not find evidence for non-zero acceleration or NUT parameters. Thus,
our simulations would suggest that the spacetime indeed approaches
Kerr, which incidentally, is also a strong test of the stability
of the Kerr solution under large, generic perturbations within the
timescales of the simulation.

These results represent the first such tests for generic binary
mergers using modest computational resources.  This naturally suggests
that further studies, perhaps also involving other numerical evolution
methods, such as Pseudo-spectral~\cite{Boyle:2006ne, Tiglio:2007jp}
and multi-patch, multi-block~\cite{Schnetter:2006pg, Zink:2007xn}, be
used to test the algebraic structure of the remnants of binary
mergers.  Finally, the algebraic structure of the remnants from the
merger of more than two black holes (e.g.\
close-encounters~\cite{Campanelli:2007ea, Lousto:2007rj} of multiple
black holes), while expected to have the same structure as the
remnants of binaries, could conceivably  have different algebraic
structures.  Thus it would be interesting to use these techniques to
examine those remnants.

\acknowledgments
We thank S.Dain for insightful comments and M.Mars for
bringing the C-metric to our attention.
We gratefully acknowledge NSF for financial support from grant PHY-0722315,
PHY-0653303, PHY 0714388, and PHY 0722703; and NASA for financial
support from grants NASA 07-ATFP07-0158 and HST-AR-11763.01.
Computational resources were provided by Lonestar cluster at TACC
and by NewHorizons at RIT.

\bibliographystyle{apsrev}
\bibliography{../../Bibtex/references}

\begin{thebibliography}{96}
\expandafter\ifx\csname natexlab\endcsname\relax\def\natexlab#1{#1}\fi
\expandafter\ifx\csname bibnamefont\endcsname\relax
  \def\bibnamefont#1{#1}\fi
\expandafter\ifx\csname bibfnamefont\endcsname\relax
  \def\bibfnamefont#1{#1}\fi
\expandafter\ifx\csname citenamefont\endcsname\relax
  \def\citenamefont#1{#1}\fi
\expandafter\ifx\csname url\endcsname\relax
  \def\url#1{\texttt{#1}}\fi
\expandafter\ifx\csname urlprefix\endcsname\relax\def\urlprefix{URL }\fi
\providecommand{\bibinfo}[2]{#2}
\providecommand{\eprint}[2][]{\url{#2}}

\bibitem[{\citenamefont{Pretorius}(2005)}]{Pretorius:2005gq}
\bibinfo{author}{\bibfnamefont{F.}~\bibnamefont{Pretorius}},
  \bibinfo{journal}{Phys. Rev. Lett.} \textbf{\bibinfo{volume}{95}},
  \bibinfo{pages}{121101} (\bibinfo{year}{2005}), \eprint{gr-qc/0507014}.

\bibitem[{\citenamefont{Campanelli
  et~al.}(2006{\natexlab{a}})\citenamefont{Campanelli, Lousto, Marronetti, and
  Zlochower}}]{Campanelli:2005dd}
\bibinfo{author}{\bibfnamefont{M.}~\bibnamefont{Campanelli}},
  \bibinfo{author}{\bibfnamefont{C.~O.} \bibnamefont{Lousto}},
  \bibinfo{author}{\bibfnamefont{P.}~\bibnamefont{Marronetti}},
  \bibnamefont{and}
  \bibinfo{author}{\bibfnamefont{Y.}~\bibnamefont{Zlochower}},
  \bibinfo{journal}{Phys. Rev. Lett.} \textbf{\bibinfo{volume}{96}},
  \bibinfo{pages}{111101} (\bibinfo{year}{2006}{\natexlab{a}}),
  \eprint{gr-qc/0511048}.

\bibitem[{\citenamefont{Baker et~al.}(2006{\natexlab{a}})\citenamefont{Baker,
  Centrella, Choi, Koppitz, and van Meter}}]{Baker:2005vv}
\bibinfo{author}{\bibfnamefont{J.~G.} \bibnamefont{Baker}},
  \bibinfo{author}{\bibfnamefont{J.}~\bibnamefont{Centrella}},
  \bibinfo{author}{\bibfnamefont{D.-I.} \bibnamefont{Choi}},
  \bibinfo{author}{\bibfnamefont{M.}~\bibnamefont{Koppitz}}, \bibnamefont{and}
  \bibinfo{author}{\bibfnamefont{J.}~\bibnamefont{van Meter}},
  \bibinfo{journal}{Phys. Rev. Lett.} \textbf{\bibinfo{volume}{96}},
  \bibinfo{pages}{111102} (\bibinfo{year}{2006}{\natexlab{a}}),
  \eprint{gr-qc/0511103}.

\bibitem[{\citenamefont{Campanelli
  et~al.}(2006{\natexlab{b}})\citenamefont{Campanelli, Lousto, and
  Zlochower}}]{Campanelli:2006uy}
\bibinfo{author}{\bibfnamefont{M.}~\bibnamefont{Campanelli}},
  \bibinfo{author}{\bibfnamefont{C.~O.} \bibnamefont{Lousto}},
  \bibnamefont{and}
  \bibinfo{author}{\bibfnamefont{Y.}~\bibnamefont{Zlochower}},
  \bibinfo{journal}{Phys. Rev. D} \textbf{\bibinfo{volume}{74}},
  \bibinfo{pages}{041501(R)} (\bibinfo{year}{2006}{\natexlab{b}}),
  \eprint{gr-qc/0604012}.

\bibitem[{\citenamefont{Campanelli
  et~al.}(2006{\natexlab{c}})\citenamefont{Campanelli, Lousto, and
  Zlochower}}]{Campanelli:2006fg}
\bibinfo{author}{\bibfnamefont{M.}~\bibnamefont{Campanelli}},
  \bibinfo{author}{\bibfnamefont{C.~O.} \bibnamefont{Lousto}},
  \bibnamefont{and}
  \bibinfo{author}{\bibfnamefont{Y.}~\bibnamefont{Zlochower}},
  \bibinfo{journal}{Phys. Rev. D} \textbf{\bibinfo{volume}{74}},
  \bibinfo{pages}{084023} (\bibinfo{year}{2006}{\natexlab{c}}),
  \eprint{astro-ph/0608275}.

\bibitem[{\citenamefont{Campanelli
  et~al.}(2007{\natexlab{a}})\citenamefont{Campanelli, Lousto, Zlochower,
  Krishnan, and Merritt}}]{Campanelli:2006fy}
\bibinfo{author}{\bibfnamefont{M.}~\bibnamefont{Campanelli}},
  \bibinfo{author}{\bibfnamefont{C.~O.} \bibnamefont{Lousto}},
  \bibinfo{author}{\bibfnamefont{Y.}~\bibnamefont{Zlochower}},
  \bibinfo{author}{\bibfnamefont{B.}~\bibnamefont{Krishnan}}, \bibnamefont{and}
  \bibinfo{author}{\bibfnamefont{D.}~\bibnamefont{Merritt}},
  \bibinfo{journal}{Phys. Rev.} \textbf{\bibinfo{volume}{D75}},
  \bibinfo{pages}{064030} (\bibinfo{year}{2007}{\natexlab{a}}),
  \eprint{gr-qc/0612076}.

\bibitem[{\citenamefont{Campanelli
  et~al.}(2006{\natexlab{d}})\citenamefont{Campanelli, Lousto, and
  Zlochower}}]{Campanelli:2006gf}
\bibinfo{author}{\bibfnamefont{M.}~\bibnamefont{Campanelli}},
  \bibinfo{author}{\bibfnamefont{C.~O.} \bibnamefont{Lousto}},
  \bibnamefont{and}
  \bibinfo{author}{\bibfnamefont{Y.}~\bibnamefont{Zlochower}},
  \bibinfo{journal}{Phys. Rev. D} \textbf{\bibinfo{volume}{73}},
  \bibinfo{pages}{061501(R)} (\bibinfo{year}{2006}{\natexlab{d}}).

\bibitem[{\citenamefont{Brugmann
  et~al.}(2008{\natexlab{a}})}]{Bruegmann:2006at}
\bibinfo{author}{\bibfnamefont{B.}~\bibnamefont{Brugmann}}
  \bibnamefont{et~al.}, \bibinfo{journal}{Phys. Rev.}
  \textbf{\bibinfo{volume}{D77}}, \bibinfo{pages}{024027}
  (\bibinfo{year}{2008}{\natexlab{a}}), \eprint{gr-qc/0610128}.

\bibitem[{\citenamefont{Baker et~al.}(2006{\natexlab{b}})\citenamefont{Baker,
  Centrella, Choi, Koppitz, and van Meter}}]{Baker:2006yw}
\bibinfo{author}{\bibfnamefont{J.~G.} \bibnamefont{Baker}},
  \bibinfo{author}{\bibfnamefont{J.}~\bibnamefont{Centrella}},
  \bibinfo{author}{\bibfnamefont{D.-I.} \bibnamefont{Choi}},
  \bibinfo{author}{\bibfnamefont{M.}~\bibnamefont{Koppitz}}, \bibnamefont{and}
  \bibinfo{author}{\bibfnamefont{J.}~\bibnamefont{van Meter}},
  \bibinfo{journal}{Phys. Rev. D} \textbf{\bibinfo{volume}{73}},
  \bibinfo{pages}{104002} (\bibinfo{year}{2006}{\natexlab{b}}),
  \eprint{gr-qc/0602026}.

\bibitem[{\citenamefont{Pretorius}(2006)}]{Pretorius:2006tp}
\bibinfo{author}{\bibfnamefont{F.}~\bibnamefont{Pretorius}},
  \bibinfo{journal}{Class. Quant. Grav.} \textbf{\bibinfo{volume}{23}},
  \bibinfo{pages}{S529} (\bibinfo{year}{2006}), \eprint{gr-qc/0602115}.

\bibitem[{\citenamefont{Pretorius and Khurana}(2007)}]{Pretorius:2007jn}
\bibinfo{author}{\bibfnamefont{F.}~\bibnamefont{Pretorius}} \bibnamefont{and}
  \bibinfo{author}{\bibfnamefont{D.}~\bibnamefont{Khurana}},
  \bibinfo{journal}{Class. Quant. Grav.} \textbf{\bibinfo{volume}{24}},
  \bibinfo{pages}{S83} (\bibinfo{year}{2007}), \eprint{gr-qc/0702084}.

\bibitem[{\citenamefont{Baker et~al.}(2007{\natexlab{a}})\citenamefont{Baker,
  van Meter, McWilliams, Centrella, and Kelly}}]{Baker:2006ha}
\bibinfo{author}{\bibfnamefont{J.~G.} \bibnamefont{Baker}},
  \bibinfo{author}{\bibfnamefont{J.~R.} \bibnamefont{van Meter}},
  \bibinfo{author}{\bibfnamefont{S.~T.} \bibnamefont{McWilliams}},
  \bibinfo{author}{\bibfnamefont{J.}~\bibnamefont{Centrella}},
  \bibnamefont{and} \bibinfo{author}{\bibfnamefont{B.~J.} \bibnamefont{Kelly}},
  \bibinfo{journal}{Phys. Rev. Lett.} \textbf{\bibinfo{volume}{99}},
  \bibinfo{pages}{181101} (\bibinfo{year}{2007}{\natexlab{a}}),
  \eprint{gr-qc/0612024}.

\bibitem[{\citenamefont{Buonanno et~al.}(2007)\citenamefont{Buonanno, Cook, and
  Pretorius}}]{Buonanno:2006ui}
\bibinfo{author}{\bibfnamefont{A.}~\bibnamefont{Buonanno}},
  \bibinfo{author}{\bibfnamefont{G.~B.} \bibnamefont{Cook}}, \bibnamefont{and}
  \bibinfo{author}{\bibfnamefont{F.}~\bibnamefont{Pretorius}},
  \bibinfo{journal}{Phys. Rev.} \textbf{\bibinfo{volume}{D75}},
  \bibinfo{pages}{124018} (\bibinfo{year}{2007}), \eprint{gr-qc/0610122}.

\bibitem[{\citenamefont{Baker et~al.}(2007{\natexlab{b}})}]{Baker:2006kr}
\bibinfo{author}{\bibfnamefont{J.~G.} \bibnamefont{Baker}}
  \bibnamefont{et~al.}, \bibinfo{journal}{Phys. Rev.}
  \textbf{\bibinfo{volume}{D75}}, \bibinfo{pages}{124024}
  (\bibinfo{year}{2007}{\natexlab{b}}), \eprint{gr-qc/0612117}.

\bibitem[{\citenamefont{Scheel et~al.}(2006)}]{Scheel:2006gg}
\bibinfo{author}{\bibfnamefont{M.~A.} \bibnamefont{Scheel}}
  \bibnamefont{et~al.}, \bibinfo{journal}{Phys. Rev.}
  \textbf{\bibinfo{volume}{D74}}, \bibinfo{pages}{104006}
  (\bibinfo{year}{2006}), \eprint{gr-qc/0607056}.

\bibitem[{\citenamefont{Baker et~al.}(2007{\natexlab{c}})\citenamefont{Baker,
  Campanelli, Pretorius, and Zlochower}}]{Baker:2007fb}
\bibinfo{author}{\bibfnamefont{J.~G.} \bibnamefont{Baker}},
  \bibinfo{author}{\bibfnamefont{M.}~\bibnamefont{Campanelli}},
  \bibinfo{author}{\bibfnamefont{F.}~\bibnamefont{Pretorius}},
  \bibnamefont{and}
  \bibinfo{author}{\bibfnamefont{Y.}~\bibnamefont{Zlochower}},
  \bibinfo{journal}{Class. Quant. Grav.} \textbf{\bibinfo{volume}{24}},
  \bibinfo{pages}{S25} (\bibinfo{year}{2007}{\natexlab{c}}),
  \eprint{gr-qc/0701016}.

\bibitem[{\citenamefont{Marronetti et~al.}(2007)}]{Marronetti:2007ya}
\bibinfo{author}{\bibfnamefont{P.}~\bibnamefont{Marronetti}}
  \bibnamefont{et~al.}, \bibinfo{journal}{Class. Quant. Grav.}
  \textbf{\bibinfo{volume}{24}}, \bibinfo{pages}{S43} (\bibinfo{year}{2007}),
  \eprint{gr-qc/0701123}.

\bibitem[{\citenamefont{Pfeiffer et~al.}(2007)}]{Pfeiffer:2007yz}
\bibinfo{author}{\bibfnamefont{H.~P.} \bibnamefont{Pfeiffer}}
  \bibnamefont{et~al.}, \bibinfo{journal}{Class. Quant. Grav.}
  \textbf{\bibinfo{volume}{24}}, \bibinfo{pages}{S59} (\bibinfo{year}{2007}),
  \eprint{gr-qc/0702106}.

\bibitem[{\citenamefont{Sperhake et~al.}(2008)\citenamefont{Sperhake, Cardoso,
  Pretorius, Berti, and Gonzalez}}]{Sperhake:2008ga}
\bibinfo{author}{\bibfnamefont{U.}~\bibnamefont{Sperhake}},
  \bibinfo{author}{\bibfnamefont{V.}~\bibnamefont{Cardoso}},
  \bibinfo{author}{\bibfnamefont{F.}~\bibnamefont{Pretorius}},
  \bibinfo{author}{\bibfnamefont{E.}~\bibnamefont{Berti}}, \bibnamefont{and}
  \bibinfo{author}{\bibfnamefont{J.~A.} \bibnamefont{Gonzalez}},
  \bibinfo{journal}{Phys. Rev. Lett.} \textbf{\bibinfo{volume}{101}},
  \bibinfo{pages}{161101} (\bibinfo{year}{2008}), \eprint{0806.1738}.

\bibitem[{\citenamefont{Campanelli
  et~al.}(2007{\natexlab{b}})\citenamefont{Campanelli, Lousto, Zlochower, and
  Merritt}}]{Campanelli:2007ew}
\bibinfo{author}{\bibfnamefont{M.}~\bibnamefont{Campanelli}},
  \bibinfo{author}{\bibfnamefont{C.~O.} \bibnamefont{Lousto}},
  \bibinfo{author}{\bibfnamefont{Y.}~\bibnamefont{Zlochower}},
  \bibnamefont{and} \bibinfo{author}{\bibfnamefont{D.}~\bibnamefont{Merritt}},
  \bibinfo{journal}{Astrophys. J.} \textbf{\bibinfo{volume}{659}},
  \bibinfo{pages}{L5} (\bibinfo{year}{2007}{\natexlab{b}}),
  \eprint{gr-qc/0701164}.

\bibitem[{\citenamefont{Campanelli
  et~al.}(2007{\natexlab{c}})\citenamefont{Campanelli, Lousto, Zlochower, and
  Merritt}}]{Campanelli:2007cga}
\bibinfo{author}{\bibfnamefont{M.}~\bibnamefont{Campanelli}},
  \bibinfo{author}{\bibfnamefont{C.~O.} \bibnamefont{Lousto}},
  \bibinfo{author}{\bibfnamefont{Y.}~\bibnamefont{Zlochower}},
  \bibnamefont{and} \bibinfo{author}{\bibfnamefont{D.}~\bibnamefont{Merritt}},
  \bibinfo{journal}{Phys. Rev. Lett.} \textbf{\bibinfo{volume}{98}},
  \bibinfo{pages}{231102} (\bibinfo{year}{2007}{\natexlab{c}}),
  \eprint{gr-qc/0702133}.

\bibitem[{\citenamefont{Krishnan et~al.}(2007)\citenamefont{Krishnan, Lousto,
  and Zlochower}}]{Krishnan:2007pu}
\bibinfo{author}{\bibfnamefont{B.}~\bibnamefont{Krishnan}},
  \bibinfo{author}{\bibfnamefont{C.~O.} \bibnamefont{Lousto}},
  \bibnamefont{and}
  \bibinfo{author}{\bibfnamefont{Y.}~\bibnamefont{Zlochower}},
  \bibinfo{journal}{Phys. Rev.} \textbf{\bibinfo{volume}{D76}},
  \bibinfo{pages}{081501} (\bibinfo{year}{2007}), \eprint{0707.0876}.

\bibitem[{\citenamefont{Dain et~al.}(2008)\citenamefont{Dain, Lousto, and
  Zlochower}}]{Dain:2008ck}
\bibinfo{author}{\bibfnamefont{S.}~\bibnamefont{Dain}},
  \bibinfo{author}{\bibfnamefont{C.~O.} \bibnamefont{Lousto}},
  \bibnamefont{and}
  \bibinfo{author}{\bibfnamefont{Y.}~\bibnamefont{Zlochower}},
  \bibinfo{journal}{Phys. Rev. D} \textbf{\bibinfo{volume}{78}},
  \bibinfo{pages}{024039} (\bibinfo{year}{2008}), \eprint{0803.0351}.

\bibitem[{\citenamefont{Campanelli}(2005)}]{Campanelli:2004zw}
\bibinfo{author}{\bibfnamefont{M.}~\bibnamefont{Campanelli}},
  \bibinfo{journal}{Class. Quant. Grav.} \textbf{\bibinfo{volume}{22}},
  \bibinfo{pages}{S387} (\bibinfo{year}{2005}), \eprint{astro-ph/0411744}.

\bibitem[{\citenamefont{Herrmann
  et~al.}(2006{\natexlab{a}})\citenamefont{Herrmann, Shoemaker, and
  Laguna}}]{Herrmann:2006ks}
\bibinfo{author}{\bibfnamefont{F.}~\bibnamefont{Herrmann}},
  \bibinfo{author}{\bibfnamefont{D.}~\bibnamefont{Shoemaker}},
  \bibnamefont{and} \bibinfo{author}{\bibfnamefont{P.}~\bibnamefont{Laguna}},
  \bibinfo{journal}{AIP Conf.} \textbf{\bibinfo{volume}{873}},
  \bibinfo{pages}{89} (\bibinfo{year}{2006}{\natexlab{a}}),
  \eprint{gr-qc/0601026}.

\bibitem[{\citenamefont{Baker et~al.}(2006{\natexlab{c}})}]{Baker:2006vn}
\bibinfo{author}{\bibfnamefont{J.~G.} \bibnamefont{Baker}}
  \bibnamefont{et~al.}, \bibinfo{journal}{Astrophys. J.}
  \textbf{\bibinfo{volume}{653}}, \bibinfo{pages}{L93}
  (\bibinfo{year}{2006}{\natexlab{c}}), \eprint{astro-ph/0603204}.

\bibitem[{\citenamefont{Sopuerta et~al.}(2006)\citenamefont{Sopuerta, Yunes,
  and Laguna}}]{Sopuerta:2006wj}
\bibinfo{author}{\bibfnamefont{C.~F.} \bibnamefont{Sopuerta}},
  \bibinfo{author}{\bibfnamefont{N.}~\bibnamefont{Yunes}}, \bibnamefont{and}
  \bibinfo{author}{\bibfnamefont{P.}~\bibnamefont{Laguna}},
  \bibinfo{journal}{Phys. Rev. D} \textbf{\bibinfo{volume}{74}},
  \bibinfo{pages}{124010} (\bibinfo{year}{2006}), \eprint{astro-ph/0608600}.

\bibitem[{\citenamefont{Gonz\'alez
  et~al.}(2007{\natexlab{a}})\citenamefont{Gonz\'alez, Sperhake, Brugmann,
  Hannam, and Husa}}]{Gonzalez:2006md}
\bibinfo{author}{\bibfnamefont{J.~A.} \bibnamefont{Gonz\'alez}},
  \bibinfo{author}{\bibfnamefont{U.}~\bibnamefont{Sperhake}},
  \bibinfo{author}{\bibfnamefont{B.}~\bibnamefont{Brugmann}},
  \bibinfo{author}{\bibfnamefont{M.}~\bibnamefont{Hannam}}, \bibnamefont{and}
  \bibinfo{author}{\bibfnamefont{S.}~\bibnamefont{Husa}},
  \bibinfo{journal}{Phys. Rev. Lett.} \textbf{\bibinfo{volume}{98}},
  \bibinfo{pages}{091101} (\bibinfo{year}{2007}{\natexlab{a}}),
  \eprint{gr-qc/0610154}.

\bibitem[{\citenamefont{Sopuerta et~al.}(2007)\citenamefont{Sopuerta, Yunes,
  and Laguna}}]{Sopuerta:2006et}
\bibinfo{author}{\bibfnamefont{C.~F.} \bibnamefont{Sopuerta}},
  \bibinfo{author}{\bibfnamefont{N.}~\bibnamefont{Yunes}}, \bibnamefont{and}
  \bibinfo{author}{\bibfnamefont{P.}~\bibnamefont{Laguna}},
  \bibinfo{journal}{Astrophys. J.} \textbf{\bibinfo{volume}{656}},
  \bibinfo{pages}{L9} (\bibinfo{year}{2007}), \eprint{astro-ph/0611110}.

\bibitem[{\citenamefont{Herrmann
  et~al.}(2006{\natexlab{b}})\citenamefont{Herrmann, Hinder, Shoemaker, and
  Laguna}}]{Herrmann:2006cd}
\bibinfo{author}{\bibfnamefont{F.}~\bibnamefont{Herrmann}},
  \bibinfo{author}{\bibfnamefont{I.}~\bibnamefont{Hinder}},
  \bibinfo{author}{\bibfnamefont{D.}~\bibnamefont{Shoemaker}},
  \bibnamefont{and} \bibinfo{author}{\bibfnamefont{P.}~\bibnamefont{Laguna}},
  \bibinfo{journal}{AIP Conf. Proc.} \textbf{\bibinfo{volume}{873}},
  \bibinfo{pages}{89} (\bibinfo{year}{2006}{\natexlab{b}}).

\bibitem[{\citenamefont{Herrmann
  et~al.}(2007{\natexlab{a}})\citenamefont{Herrmann, Hinder, Shoemaker, and
  Laguna}}]{Herrmann:2007zz}
\bibinfo{author}{\bibfnamefont{F.}~\bibnamefont{Herrmann}},
  \bibinfo{author}{\bibfnamefont{I.}~\bibnamefont{Hinder}},
  \bibinfo{author}{\bibfnamefont{D.}~\bibnamefont{Shoemaker}},
  \bibnamefont{and} \bibinfo{author}{\bibfnamefont{P.}~\bibnamefont{Laguna}},
  \bibinfo{journal}{Class. Quant. Grav.} \textbf{\bibinfo{volume}{24}},
  \bibinfo{pages}{S33} (\bibinfo{year}{2007}{\natexlab{a}}).

\bibitem[{\citenamefont{Herrmann
  et~al.}(2007{\natexlab{b}})\citenamefont{Herrmann, Hinder, Shoemaker, Laguna,
  and Matzner}}]{Herrmann:2007ac}
\bibinfo{author}{\bibfnamefont{F.}~\bibnamefont{Herrmann}},
  \bibinfo{author}{\bibfnamefont{I.}~\bibnamefont{Hinder}},
  \bibinfo{author}{\bibfnamefont{D.}~\bibnamefont{Shoemaker}},
  \bibinfo{author}{\bibfnamefont{P.}~\bibnamefont{Laguna}}, \bibnamefont{and}
  \bibinfo{author}{\bibfnamefont{R.~A.} \bibnamefont{Matzner}},
  \bibinfo{journal}{Astrophys. J.} \textbf{\bibinfo{volume}{661}},
  \bibinfo{pages}{430} (\bibinfo{year}{2007}{\natexlab{b}}),
  \eprint{gr-qc/0701143}.

\bibitem[{\citenamefont{Koppitz et~al.}(2007)}]{Koppitz:2007ev}
\bibinfo{author}{\bibfnamefont{M.}~\bibnamefont{Koppitz}} \bibnamefont{et~al.},
  \bibinfo{journal}{Phys. Rev. Lett.} \textbf{\bibinfo{volume}{99}},
  \bibinfo{pages}{041102} (\bibinfo{year}{2007}), \eprint{gr-qc/0701163}.

\bibitem[{\citenamefont{Choi et~al.}(2007)}]{Choi:2007eu}
\bibinfo{author}{\bibfnamefont{D.-I.} \bibnamefont{Choi}} \bibnamefont{et~al.},
  \bibinfo{journal}{Phys. Rev.} \textbf{\bibinfo{volume}{D76}},
  \bibinfo{pages}{104026} (\bibinfo{year}{2007}), \eprint{gr-qc/0702016}.

\bibitem[{\citenamefont{Gonz\'alez
  et~al.}(2007{\natexlab{b}})\citenamefont{Gonz\'alez, Hannam, Sperhake,
  Brugmann, and Husa}}]{Gonzalez:2007hi}
\bibinfo{author}{\bibfnamefont{J.~A.} \bibnamefont{Gonz\'alez}},
  \bibinfo{author}{\bibfnamefont{M.~D.} \bibnamefont{Hannam}},
  \bibinfo{author}{\bibfnamefont{U.}~\bibnamefont{Sperhake}},
  \bibinfo{author}{\bibfnamefont{B.}~\bibnamefont{Brugmann}}, \bibnamefont{and}
  \bibinfo{author}{\bibfnamefont{S.}~\bibnamefont{Husa}},
  \bibinfo{journal}{Phys. Rev. Lett.} \textbf{\bibinfo{volume}{98}},
  \bibinfo{pages}{231101} (\bibinfo{year}{2007}{\natexlab{b}}),
  \eprint{gr-qc/0702052}.

\bibitem[{\citenamefont{Baker et~al.}(2007{\natexlab{d}})}]{Baker:2007gi}
\bibinfo{author}{\bibfnamefont{J.~G.} \bibnamefont{Baker}}
  \bibnamefont{et~al.}, \bibinfo{journal}{Astrophys. J.}
  \textbf{\bibinfo{volume}{668}}, \bibinfo{pages}{1140}
  (\bibinfo{year}{2007}{\natexlab{d}}), \eprint{astro-ph/0702390}.

\bibitem[{\citenamefont{Berti et~al.}(2007)}]{Berti:2007fi}
\bibinfo{author}{\bibfnamefont{E.}~\bibnamefont{Berti}} \bibnamefont{et~al.},
  \bibinfo{journal}{Phys. Rev.} \textbf{\bibinfo{volume}{D76}},
  \bibinfo{pages}{064034} (\bibinfo{year}{2007}), \eprint{gr-qc/0703053}.

\bibitem[{\citenamefont{Tichy and Marronetti}(2007)}]{Tichy:2007hk}
\bibinfo{author}{\bibfnamefont{W.}~\bibnamefont{Tichy}} \bibnamefont{and}
  \bibinfo{author}{\bibfnamefont{P.}~\bibnamefont{Marronetti}},
  \bibinfo{journal}{Phys. Rev.} \textbf{\bibinfo{volume}{D76}},
  \bibinfo{pages}{061502} (\bibinfo{year}{2007}), \eprint{gr-qc/0703075}.

\bibitem[{\citenamefont{Herrmann
  et~al.}(2007{\natexlab{c}})\citenamefont{Herrmann, Hinder, Shoemaker, Laguna,
  and Matzner}}]{Herrmann:2007ex}
\bibinfo{author}{\bibfnamefont{F.}~\bibnamefont{Herrmann}},
  \bibinfo{author}{\bibfnamefont{I.}~\bibnamefont{Hinder}},
  \bibinfo{author}{\bibfnamefont{D.~M.} \bibnamefont{Shoemaker}},
  \bibinfo{author}{\bibfnamefont{P.}~\bibnamefont{Laguna}}, \bibnamefont{and}
  \bibinfo{author}{\bibfnamefont{R.~A.} \bibnamefont{Matzner}},
  \bibinfo{journal}{Phys. Rev.} \textbf{\bibinfo{volume}{D76}},
  \bibinfo{pages}{084032} (\bibinfo{year}{2007}{\natexlab{c}}),
  \eprint{0706.2541}.

\bibitem[{\citenamefont{Brugmann
  et~al.}(2008{\natexlab{b}})\citenamefont{Brugmann, Gonzalez, Hannam, Husa,
  and Sperhake}}]{Brugmann:2007zj}
\bibinfo{author}{\bibfnamefont{B.}~\bibnamefont{Brugmann}},
  \bibinfo{author}{\bibfnamefont{J.~A.} \bibnamefont{Gonzalez}},
  \bibinfo{author}{\bibfnamefont{M.}~\bibnamefont{Hannam}},
  \bibinfo{author}{\bibfnamefont{S.}~\bibnamefont{Husa}}, \bibnamefont{and}
  \bibinfo{author}{\bibfnamefont{U.}~\bibnamefont{Sperhake}},
  \bibinfo{journal}{Phys. Rev.} \textbf{\bibinfo{volume}{D77}},
  \bibinfo{pages}{124047} (\bibinfo{year}{2008}{\natexlab{b}}),
  \eprint{0707.0135}.

\bibitem[{\citenamefont{Schnittman et~al.}(2008)}]{Schnittman:2007ij}
\bibinfo{author}{\bibfnamefont{J.~D.} \bibnamefont{Schnittman}}
  \bibnamefont{et~al.}, \bibinfo{journal}{Phys. Rev.}
  \textbf{\bibinfo{volume}{D77}}, \bibinfo{pages}{044031}
  (\bibinfo{year}{2008}), \eprint{0707.0301}.

\bibitem[{\citenamefont{Holley-Bockelmann
  et~al.}(2007)\citenamefont{Holley-Bockelmann, Gultekin, Shoemaker, and
  Yunes}}]{HolleyBockelmann:2007eh}
\bibinfo{author}{\bibfnamefont{K.}~\bibnamefont{Holley-Bockelmann}},
  \bibinfo{author}{\bibfnamefont{K.}~\bibnamefont{Gultekin}},
  \bibinfo{author}{\bibfnamefont{D.}~\bibnamefont{Shoemaker}},
  \bibnamefont{and} \bibinfo{author}{\bibfnamefont{N.}~\bibnamefont{Yunes}}
  (\bibinfo{year}{2007}), \eprint{0707.1334}.

\bibitem[{\citenamefont{Pollney et~al.}(2007)}]{Pollney:2007ss}
\bibinfo{author}{\bibfnamefont{D.}~\bibnamefont{Pollney}} \bibnamefont{et~al.},
  \bibinfo{journal}{Phys. Rev.} \textbf{\bibinfo{volume}{D76}},
  \bibinfo{pages}{124002} (\bibinfo{year}{2007}), \eprint{0707.2559}.

\bibitem[{\citenamefont{{Redmount} and {Rees}}(1989)}]{Redmount:1989}
\bibinfo{author}{\bibfnamefont{I.~H.} \bibnamefont{{Redmount}}}
  \bibnamefont{and} \bibinfo{author}{\bibfnamefont{M.~J.}
  \bibnamefont{{Rees}}}, \bibinfo{journal}{Comments on Astrophysics}
  \textbf{\bibinfo{volume}{14}}, \bibinfo{pages}{165} (\bibinfo{year}{1989}).

\bibitem[{\citenamefont{Merritt et~al.}(2004)\citenamefont{Merritt,
  Milosavljevic, Favata, Hughes, and Holz}}]{Merritt:2004xa}
\bibinfo{author}{\bibfnamefont{D.}~\bibnamefont{Merritt}},
  \bibinfo{author}{\bibfnamefont{M.}~\bibnamefont{Milosavljevic}},
  \bibinfo{author}{\bibfnamefont{M.}~\bibnamefont{Favata}},
  \bibinfo{author}{\bibfnamefont{S.~A.} \bibnamefont{Hughes}},
  \bibnamefont{and} \bibinfo{author}{\bibfnamefont{D.~E.} \bibnamefont{Holz}},
  \bibinfo{journal}{Astrophys. J.} \textbf{\bibinfo{volume}{607}},
  \bibinfo{pages}{L9} (\bibinfo{year}{2004}), \eprint{astro-ph/0402057}.

\bibitem[{\citenamefont{Gualandris and Merritt}(2007)}]{Gualandris:2007nm}
\bibinfo{author}{\bibfnamefont{A.}~\bibnamefont{Gualandris}} \bibnamefont{and}
  \bibinfo{author}{\bibfnamefont{D.}~\bibnamefont{Merritt}}
  (\bibinfo{year}{2007}), \eprint{0708.0771}.

\bibitem[{\citenamefont{Kapoor}(1976)}]{Kapoor76}
\bibinfo{author}{\bibfnamefont{R.~C.} \bibnamefont{Kapoor}},
  \bibinfo{journal}{Pramana} \textbf{\bibinfo{volume}{7}}, \bibinfo{pages}{334}
  (\bibinfo{year}{1976}).

\bibitem[{\citenamefont{Bogdanovic et~al.}(2007)\citenamefont{Bogdanovic,
  Reynolds, and Miller}}]{Bogdanovic:2007hp}
\bibinfo{author}{\bibfnamefont{T.}~\bibnamefont{Bogdanovic}},
  \bibinfo{author}{\bibfnamefont{C.~S.} \bibnamefont{Reynolds}},
  \bibnamefont{and} \bibinfo{author}{\bibfnamefont{M.~C.} \bibnamefont{Miller}}
  (\bibinfo{year}{2007}), \eprint{astro-ph/0703054}.

\bibitem[{\citenamefont{Loeb}(2007)}]{Loeb:2007wz}
\bibinfo{author}{\bibfnamefont{A.}~\bibnamefont{Loeb}}, \bibinfo{journal}{Phys.
  Rev. Lett.} \textbf{\bibinfo{volume}{99}}, \bibinfo{pages}{041103}
  (\bibinfo{year}{2007}), \eprint{astro-ph/0703722}.

\bibitem[{\citenamefont{Bonning et~al.}(2007)\citenamefont{Bonning, Shields,
  and Salviander}}]{Bonning:2007vt}
\bibinfo{author}{\bibfnamefont{E.~W.} \bibnamefont{Bonning}},
  \bibinfo{author}{\bibfnamefont{G.~A.} \bibnamefont{Shields}},
  \bibnamefont{and}
  \bibinfo{author}{\bibfnamefont{S.}~\bibnamefont{Salviander}}
  (\bibinfo{year}{2007}), \eprint{0705.4263}.

\bibitem[{\citenamefont{Komossa et~al.}(2008)\citenamefont{Komossa, Zhou, and
  Lu}}]{Komossa:2008qd}
\bibinfo{author}{\bibfnamefont{S.}~\bibnamefont{Komossa}},
  \bibinfo{author}{\bibfnamefont{H.}~\bibnamefont{Zhou}}, \bibnamefont{and}
  \bibinfo{author}{\bibfnamefont{H.}~\bibnamefont{Lu}},
  \bibinfo{journal}{Astrop. J. Letters} \textbf{\bibinfo{volume}{678}},
  \bibinfo{pages}{L81} (\bibinfo{year}{2008}), \eprint{0804.4585}.

\bibitem[{\citenamefont{Komossa and Merritt}(2008)}]{Komossa:2008ye}
\bibinfo{author}{\bibfnamefont{S.}~\bibnamefont{Komossa}} \bibnamefont{and}
  \bibinfo{author}{\bibfnamefont{D.}~\bibnamefont{Merritt}},
  \bibinfo{journal}{Astrophys. J.} \textbf{\bibinfo{volume}{683}},
  \bibinfo{pages}{L21} (\bibinfo{year}{2008}), \eprint{0807.0223}.

\bibitem[{\citenamefont{Shields et~al.}(2008)\citenamefont{Shields, Bonning,
  and Salviander}}]{Shields:2008kn}
\bibinfo{author}{\bibfnamefont{G.~A.} \bibnamefont{Shields}},
  \bibinfo{author}{\bibfnamefont{E.~W.} \bibnamefont{Bonning}},
  \bibnamefont{and}
  \bibinfo{author}{\bibfnamefont{S.}~\bibnamefont{Salviander}}
  (\bibinfo{year}{2008}), \eprint{0810.2563}.

\bibitem[{\citenamefont{Rezzolla et~al.}(2008)}]{Rezzolla:2007rd}
\bibinfo{author}{\bibfnamefont{L.}~\bibnamefont{Rezzolla}}
  \bibnamefont{et~al.}, \bibinfo{journal}{Astrophys. J.}
  \textbf{\bibinfo{volume}{674}}, \bibinfo{pages}{L29} (\bibinfo{year}{2008}),
  \eprint{arXiv:0710.3345 [gr-qc]}.

\bibitem[{\citenamefont{Sperhake et~al.}(2007)}]{Sperhake:2007gu}
\bibinfo{author}{\bibfnamefont{U.}~\bibnamefont{Sperhake}} \bibnamefont{et~al.}
  (\bibinfo{year}{2007}), \eprint{arXiv:0710.3823 [gr-qc]}.

\bibitem[{\citenamefont{Lousto and Zlochower}(2008)}]{Lousto:2007rj}
\bibinfo{author}{\bibfnamefont{C.~O.} \bibnamefont{Lousto}} \bibnamefont{and}
  \bibinfo{author}{\bibfnamefont{Y.}~\bibnamefont{Zlochower}},
  \bibinfo{journal}{Phys. Rev.} \textbf{\bibinfo{volume}{D77}},
  \bibinfo{pages}{024034} (\bibinfo{year}{2008}), \eprint{0711.1165}.

\bibitem[{\citenamefont{Campanelli
  et~al.}(2008{\natexlab{a}})\citenamefont{Campanelli, Lousto, and
  Zlochower}}]{Campanelli:2007ea}
\bibinfo{author}{\bibfnamefont{M.}~\bibnamefont{Campanelli}},
  \bibinfo{author}{\bibfnamefont{C.~O.} \bibnamefont{Lousto}},
  \bibnamefont{and}
  \bibinfo{author}{\bibfnamefont{Y.}~\bibnamefont{Zlochower}},
  \bibinfo{journal}{Phys. Rev. D} \textbf{\bibinfo{volume}{77}},
  \bibinfo{pages}{101501(R)} (\bibinfo{year}{2008}{\natexlab{a}}),
  \eprint{0710.0879}.

\bibitem[{\citenamefont{Mazur}(2000)}]{Mazur:2000pn}
\bibinfo{author}{\bibfnamefont{P.~O.} \bibnamefont{Mazur}}
  (\bibinfo{year}{2000}), \eprint{hep-th/0101012}.

\bibitem[{\citenamefont{Campanelli
  et~al.}(2006{\natexlab{e}})\citenamefont{Campanelli, Kelly, and
  Lousto}}]{Campanelli:2005ia}
\bibinfo{author}{\bibfnamefont{M.}~\bibnamefont{Campanelli}},
  \bibinfo{author}{\bibfnamefont{B.}~\bibnamefont{Kelly}}, \bibnamefont{and}
  \bibinfo{author}{\bibfnamefont{C.~O.} \bibnamefont{Lousto}},
  \bibinfo{journal}{Phys. Rev. D} \textbf{\bibinfo{volume}{73}},
  \bibinfo{pages}{064005} (\bibinfo{year}{2006}{\natexlab{e}}),
  \eprint{gr-qc/0510122}.

\bibitem[{\citenamefont{Baker et~al.}(2002{\natexlab{a}})\citenamefont{Baker,
  Campanelli, and Lousto}}]{Baker:2001sf}
\bibinfo{author}{\bibfnamefont{J.}~\bibnamefont{Baker}},
  \bibinfo{author}{\bibfnamefont{M.}~\bibnamefont{Campanelli}},
  \bibnamefont{and} \bibinfo{author}{\bibfnamefont{C.~O.}
  \bibnamefont{Lousto}}, \bibinfo{journal}{Phys. Rev. D}
  \textbf{\bibinfo{volume}{65}}, \bibinfo{pages}{044001}
  (\bibinfo{year}{2002}{\natexlab{a}}), \eprint{gr-qc/0104063}.

\bibitem[{\citenamefont{Baker and Campanelli}(2000)}]{Baker00a}
\bibinfo{author}{\bibfnamefont{J.}~\bibnamefont{Baker}} \bibnamefont{and}
  \bibinfo{author}{\bibfnamefont{M.}~\bibnamefont{Campanelli}},
  \bibinfo{journal}{Phys. Rev. D} \textbf{\bibinfo{volume}{62}},
  \bibinfo{pages}{127501} (\bibinfo{year}{2000}).

\bibitem[{\citenamefont{Dreyer et~al.}(2003)\citenamefont{Dreyer, Krishnan,
  Shoemaker, and Schnetter}}]{Dreyer02a}
\bibinfo{author}{\bibfnamefont{O.}~\bibnamefont{Dreyer}},
  \bibinfo{author}{\bibfnamefont{B.}~\bibnamefont{Krishnan}},
  \bibinfo{author}{\bibfnamefont{D.}~\bibnamefont{Shoemaker}},
  \bibnamefont{and}
  \bibinfo{author}{\bibfnamefont{E.}~\bibnamefont{Schnetter}},
  \bibinfo{journal}{Phys. Rev. D} \textbf{\bibinfo{volume}{67}},
  \bibinfo{pages}{024018} (\bibinfo{year}{2003}), \eprint{gr-qc/0206008}.

\bibitem[{\citenamefont{Campanelli and Lousto}(1999)}]{Campanelli99}
\bibinfo{author}{\bibfnamefont{M.}~\bibnamefont{Campanelli}} \bibnamefont{and}
  \bibinfo{author}{\bibfnamefont{C.~O.} \bibnamefont{Lousto}},
  \bibinfo{journal}{Phys. Rev. D} \textbf{\bibinfo{volume}{59}},
  \bibinfo{pages}{124022} (\bibinfo{year}{1999}), \eprint{gr-qc/9811019}.

\bibitem[{\citenamefont{Lousto and Zlochower}(2007)}]{Lousto:2007mh}
\bibinfo{author}{\bibfnamefont{C.~O.} \bibnamefont{Lousto}} \bibnamefont{and}
  \bibinfo{author}{\bibfnamefont{Y.}~\bibnamefont{Zlochower}},
  \bibinfo{journal}{Phys. Rev. D} \textbf{\bibinfo{volume}{76}},
  \bibinfo{pages}{041502(R)} (\bibinfo{year}{2007}), \eprint{gr-qc/0703061}.

\bibitem[{\citenamefont{Scheel et~al.}(2009)}]{Scheel:2008rj}
\bibinfo{author}{\bibfnamefont{M.~A.} \bibnamefont{Scheel}}
  \bibnamefont{et~al.}, \bibinfo{journal}{Phys. Rev.}
  \textbf{\bibinfo{volume}{D79}}, \bibinfo{pages}{024003}
  (\bibinfo{year}{2009}), \eprint{0810.1767}.

\bibitem[{\citenamefont{Mars}(1999)}]{Mars:1999yn}
\bibinfo{author}{\bibfnamefont{M.}~\bibnamefont{Mars}},
  \bibinfo{journal}{Class. Quant. Grav.} \textbf{\bibinfo{volume}{16}},
  \bibinfo{pages}{2507} (\bibinfo{year}{1999}), \eprint{gr-qc/9904070}.

\bibitem[{\citenamefont{Mars}(2000)}]{Mars:2000gb}
\bibinfo{author}{\bibfnamefont{M.}~\bibnamefont{Mars}},
  \bibinfo{journal}{Class. Quant. Grav.} \textbf{\bibinfo{volume}{17}},
  \bibinfo{pages}{3353} (\bibinfo{year}{2000}), \eprint{gr-qc/0004018}.

\bibitem[{\citenamefont{Whiting}(1989)}]{Whiting:1989vc}
\bibinfo{author}{\bibfnamefont{B.~F.} \bibnamefont{Whiting}},
  \bibinfo{journal}{J. Math. Phys.} \textbf{\bibinfo{volume}{30}},
  \bibinfo{pages}{1301} (\bibinfo{year}{1989}).

\bibitem[{\citenamefont{Dotti et~al.}(2008)\citenamefont{Dotti, Gleiser,
  Ranea-Sandoval, and Vucetich}}]{Dotti:2008yr}
\bibinfo{author}{\bibfnamefont{G.}~\bibnamefont{Dotti}},
  \bibinfo{author}{\bibfnamefont{R.~J.} \bibnamefont{Gleiser}},
  \bibinfo{author}{\bibfnamefont{I.~F.} \bibnamefont{Ranea-Sandoval}},
  \bibnamefont{and} \bibinfo{author}{\bibfnamefont{H.}~\bibnamefont{Vucetich}}
  (\bibinfo{year}{2008}), \eprint{0805.4306}.

\bibitem[{\citenamefont{Stephani et~al.}(2003)\citenamefont{Stephani, Kramer,
  MacCallum, Hoenselaers, and Herlt}}]{Stephani:2003tm}
\bibinfo{author}{\bibfnamefont{H.}~\bibnamefont{Stephani}},
  \bibinfo{author}{\bibfnamefont{D.}~\bibnamefont{Kramer}},
  \bibinfo{author}{\bibfnamefont{M.}~\bibnamefont{MacCallum}},
  \bibinfo{author}{\bibfnamefont{C.}~\bibnamefont{Hoenselaers}},
  \bibnamefont{and} \bibinfo{author}{\bibfnamefont{E.}~\bibnamefont{Herlt}},
  \emph{\bibinfo{title}{{Exact solutions to Einstein's field equations}}}
  (\bibinfo{publisher}{Cambridge, UK: Univ. Pr.}, \bibinfo{year}{2003}),
  \bibinfo{note}{2nd Edition. 701 P.}

\bibitem[{\citenamefont{d'Inverno and Russel-Clark}(1971)}]{dInverno71}
\bibinfo{author}{\bibfnamefont{R.~A.} \bibnamefont{d'Inverno}}
  \bibnamefont{and} \bibinfo{author}{\bibfnamefont{R.~A.}
  \bibnamefont{Russel-Clark}}, \bibinfo{journal}{J. Math. Phys.}
  \textbf{\bibinfo{volume}{12}}, \bibinfo{pages}{1258} (\bibinfo{year}{1971}).

\bibitem[{\citenamefont{Carminati and McLenaghan}(1991)}]{Carminati91}
\bibinfo{author}{\bibfnamefont{J.}~\bibnamefont{Carminati}} \bibnamefont{and}
  \bibinfo{author}{\bibfnamefont{R.}~\bibnamefont{McLenaghan}},
  \bibinfo{journal}{J. Math. Phys.} \textbf{\bibinfo{volume}{32}},
  \bibinfo{pages}{3135} (\bibinfo{year}{1991}).

\bibitem[{\citenamefont{Gunnarsen et~al.}(1995)\citenamefont{Gunnarsen,
  Shinkai, and Maeda}}]{Gunnarsen95}
\bibinfo{author}{\bibfnamefont{L.}~\bibnamefont{Gunnarsen}},
  \bibinfo{author}{\bibfnamefont{H.}~\bibnamefont{Shinkai}}, \bibnamefont{and}
  \bibinfo{author}{\bibfnamefont{K.}~\bibnamefont{Maeda}},
  \bibinfo{journal}{Class. Quantum Grav.} \textbf{\bibinfo{volume}{12}},
  \bibinfo{pages}{133} (\bibinfo{year}{1995}), \eprint{gr-qc/9406003}.

\bibitem[{\citenamefont{Chandrasekhar}(1983)}]{Chandrasekhar83}
\bibinfo{author}{\bibfnamefont{S.}~\bibnamefont{Chandrasekhar}},
  \emph{\bibinfo{title}{The Mathematical Theory of Black Holes}}
  (\bibinfo{publisher}{Oxford University Press}, \bibinfo{address}{Oxford,
  England}, \bibinfo{year}{1983}).

\bibitem[{\citenamefont{Abramowitz and Stegun}(1972)}]{AbramowitzStegun}
\bibinfo{author}{\bibfnamefont{M.}~\bibnamefont{Abramowitz}} \bibnamefont{and}
  \bibinfo{author}{\bibfnamefont{I.}~\bibnamefont{Stegun}},
  \emph{\bibinfo{title}{Handbook of Mathematical Functions}}
  (\bibinfo{publisher}{Dover}, \bibinfo{address}{New York},
  \bibinfo{year}{1972}), \bibinfo{edition}{10th} ed.

\bibitem[{\citenamefont{Stephani}(2004)}]{Stephani:2004ac}
\bibinfo{author}{\bibfnamefont{H.}~\bibnamefont{Stephani}},
  \emph{\bibinfo{title}{Relativity : an introduction to special and general
  relativity}} (\bibinfo{publisher}{Cambridge Univ. Pr.},
  \bibinfo{address}{Cambridge, UK}, \bibinfo{year}{2004}),
  \bibinfo{edition}{3rd} ed.

\bibitem[{\citenamefont{Griffiths and Podolsky}(2005)}]{Griffiths:2005se}
\bibinfo{author}{\bibfnamefont{J.~B.} \bibnamefont{Griffiths}}
  \bibnamefont{and} \bibinfo{author}{\bibfnamefont{J.}~\bibnamefont{Podolsky}},
  \bibinfo{journal}{Class. Quant. Grav.} \textbf{\bibinfo{volume}{22}},
  \bibinfo{pages}{3467} (\bibinfo{year}{2005}), \eprint{gr-qc/0507021}.

\bibitem[{\citenamefont{Beetle et~al.}(2005)\citenamefont{Beetle, Bruni, Burko,
  and Nerozzi}}]{Beetle:2004wu}
\bibinfo{author}{\bibfnamefont{C.}~\bibnamefont{Beetle}},
  \bibinfo{author}{\bibfnamefont{M.}~\bibnamefont{Bruni}},
  \bibinfo{author}{\bibfnamefont{L.~M.} \bibnamefont{Burko}}, \bibnamefont{and}
  \bibinfo{author}{\bibfnamefont{A.}~\bibnamefont{Nerozzi}},
  \bibinfo{journal}{Phys. Rev. D} \textbf{\bibinfo{volume}{72}},
  \bibinfo{pages}{024013} (\bibinfo{year}{2005}), \eprint{gr-qc/0407012}.

\bibitem[{\citenamefont{Brandt and Br{\"u}gmann}(1997)}]{Brandt97b}
\bibinfo{author}{\bibfnamefont{S.}~\bibnamefont{Brandt}} \bibnamefont{and}
  \bibinfo{author}{\bibfnamefont{B.}~\bibnamefont{Br{\"u}gmann}},
  \bibinfo{journal}{Phys. Rev. Lett.} \textbf{\bibinfo{volume}{78}},
  \bibinfo{pages}{3606} (\bibinfo{year}{1997}), \eprint{gr-qc/9703066}.

\bibitem[{\citenamefont{Ansorg et~al.}(2004)\citenamefont{Ansorg, Br\"ugmann,
  and Tichy}}]{Ansorg:2004ds}
\bibinfo{author}{\bibfnamefont{M.}~\bibnamefont{Ansorg}},
  \bibinfo{author}{\bibfnamefont{B.}~\bibnamefont{Br\"ugmann}},
  \bibnamefont{and} \bibinfo{author}{\bibfnamefont{W.}~\bibnamefont{Tichy}},
  \bibinfo{journal}{Phys. Rev. D} \textbf{\bibinfo{volume}{70}},
  \bibinfo{pages}{064011} (\bibinfo{year}{2004}), \eprint{gr-qc/0404056}.

\bibitem[{\citenamefont{Zlochower et~al.}(2005)\citenamefont{Zlochower, Baker,
  Campanelli, and Lousto}}]{Zlochower:2005bj}
\bibinfo{author}{\bibfnamefont{Y.}~\bibnamefont{Zlochower}},
  \bibinfo{author}{\bibfnamefont{J.~G.} \bibnamefont{Baker}},
  \bibinfo{author}{\bibfnamefont{M.}~\bibnamefont{Campanelli}},
  \bibnamefont{and} \bibinfo{author}{\bibfnamefont{C.~O.}
  \bibnamefont{Lousto}}, \bibinfo{journal}{Phys. Rev. D}
  \textbf{\bibinfo{volume}{72}}, \bibinfo{pages}{024021}
  (\bibinfo{year}{2005}), \eprint{gr-qc/0505055}.

\bibitem[{\citenamefont{Nakamura et~al.}(1987)\citenamefont{Nakamura, Oohara,
  and Kojima}}]{Nakamura87}
\bibinfo{author}{\bibfnamefont{T.}~\bibnamefont{Nakamura}},
  \bibinfo{author}{\bibfnamefont{K.}~\bibnamefont{Oohara}}, \bibnamefont{and}
  \bibinfo{author}{\bibfnamefont{Y.}~\bibnamefont{Kojima}},
  \bibinfo{journal}{Prog. Theor. Phys. Suppl.} \textbf{\bibinfo{volume}{90}},
  \bibinfo{pages}{1} (\bibinfo{year}{1987}).

\bibitem[{\citenamefont{Shibata and Nakamura}(1995)}]{Shibata95}
\bibinfo{author}{\bibfnamefont{M.}~\bibnamefont{Shibata}} \bibnamefont{and}
  \bibinfo{author}{\bibfnamefont{T.}~\bibnamefont{Nakamura}},
  \bibinfo{journal}{Phys. Rev. D} \textbf{\bibinfo{volume}{52}},
  \bibinfo{pages}{5428} (\bibinfo{year}{1995}).

\bibitem[{\citenamefont{Baumgarte and Shapiro}(1999)}]{Baumgarte99}
\bibinfo{author}{\bibfnamefont{T.~W.} \bibnamefont{Baumgarte}}
  \bibnamefont{and} \bibinfo{author}{\bibfnamefont{S.~L.}
  \bibnamefont{Shapiro}}, \bibinfo{journal}{Phys. Rev. D}
  \textbf{\bibinfo{volume}{59}}, \bibinfo{pages}{024007}
  (\bibinfo{year}{1999}), \eprint{gr-qc/9810065}.

\bibitem[{\citenamefont{Marronetti et~al.}(2008)\citenamefont{Marronetti,
  Tichy, Brugmann, Gonzalez, and Sperhake}}]{Marronetti:2007wz}
\bibinfo{author}{\bibfnamefont{P.}~\bibnamefont{Marronetti}},
  \bibinfo{author}{\bibfnamefont{W.}~\bibnamefont{Tichy}},
  \bibinfo{author}{\bibfnamefont{B.}~\bibnamefont{Brugmann}},
  \bibinfo{author}{\bibfnamefont{J.}~\bibnamefont{Gonzalez}}, \bibnamefont{and}
  \bibinfo{author}{\bibfnamefont{U.}~\bibnamefont{Sperhake}},
  \bibinfo{journal}{Phys. Rev.} \textbf{\bibinfo{volume}{D77}},
  \bibinfo{pages}{064010} (\bibinfo{year}{2008}), \eprint{0709.2160}.

\bibitem[{\citenamefont{Schnetter et~al.}(2004)\citenamefont{Schnetter, Hawley,
  and Hawke}}]{Schnetter-etal-03b}
\bibinfo{author}{\bibfnamefont{E.}~\bibnamefont{Schnetter}},
  \bibinfo{author}{\bibfnamefont{S.~H.} \bibnamefont{Hawley}},
  \bibnamefont{and} \bibinfo{author}{\bibfnamefont{I.}~\bibnamefont{Hawke}},
  \bibinfo{journal}{Class. Quantum Grav.} \textbf{\bibinfo{volume}{21}},
  \bibinfo{pages}{1465} (\bibinfo{year}{2004}), \eprint{gr-qc/0310042}.

\bibitem[{\citenamefont{Alcubierre et~al.}(2003)\citenamefont{Alcubierre,
  Br\"ugmann, Diener, Koppitz, Pollney, Seidel, and Takahashi}}]{Alcubierre02a}
\bibinfo{author}{\bibfnamefont{M.}~\bibnamefont{Alcubierre}},
  \bibinfo{author}{\bibfnamefont{B.}~\bibnamefont{Br\"ugmann}},
  \bibinfo{author}{\bibfnamefont{P.}~\bibnamefont{Diener}},
  \bibinfo{author}{\bibfnamefont{M.}~\bibnamefont{Koppitz}},
  \bibinfo{author}{\bibfnamefont{D.}~\bibnamefont{Pollney}},
  \bibinfo{author}{\bibfnamefont{E.}~\bibnamefont{Seidel}}, \bibnamefont{and}
  \bibinfo{author}{\bibfnamefont{R.}~\bibnamefont{Takahashi}},
  \bibinfo{journal}{Phys. Rev. D} \textbf{\bibinfo{volume}{67}},
  \bibinfo{pages}{084023} (\bibinfo{year}{2003}), \eprint{gr-qc/0206072}.

\bibitem[{\citenamefont{Gundlach and Martin-Garcia}(2006)}]{Gundlach:2006tw}
\bibinfo{author}{\bibfnamefont{C.}~\bibnamefont{Gundlach}} \bibnamefont{and}
  \bibinfo{author}{\bibfnamefont{J.~M.} \bibnamefont{Martin-Garcia}},
  \bibinfo{journal}{Phys. Rev.} \textbf{\bibinfo{volume}{D74}},
  \bibinfo{pages}{024016} (\bibinfo{year}{2006}), \eprint{gr-qc/0604035}.

\bibitem[{\citenamefont{Thornburg}(2004)}]{Thornburg2003:AH-finding}
\bibinfo{author}{\bibfnamefont{J.}~\bibnamefont{Thornburg}},
  \bibinfo{journal}{Class. Quantum Grav.} \textbf{\bibinfo{volume}{21}},
  \bibinfo{pages}{743} (\bibinfo{year}{2004}), \eprint{gr-qc/0306056}.

\bibitem[{\citenamefont{Campanelli
  et~al.}(2008{\natexlab{b}})\citenamefont{Campanelli, Lousto, Nakano, and
  Zlochower}}]{Campanelli:2008nk}
\bibinfo{author}{\bibfnamefont{M.}~\bibnamefont{Campanelli}},
  \bibinfo{author}{\bibfnamefont{C.~O.} \bibnamefont{Lousto}},
  \bibinfo{author}{\bibfnamefont{H.}~\bibnamefont{Nakano}}, \bibnamefont{and}
  \bibinfo{author}{\bibfnamefont{Y.}~\bibnamefont{Zlochower}}
  (\bibinfo{year}{2008}{\natexlab{b}}), \eprint{0808.0713}.

\bibitem[{\citenamefont{Echeverr\'{\i}a}(1989)}]{Echeverria89}
\bibinfo{author}{\bibfnamefont{F.}~\bibnamefont{Echeverr\'{\i}a}},
  \bibinfo{journal}{Phys. Rev. D} \textbf{\bibinfo{volume}{40}},
  \bibinfo{pages}{3194} (\bibinfo{year}{1989}).

\bibitem[{\citenamefont{Baker et~al.}(2002{\natexlab{b}})\citenamefont{Baker,
  Campanelli, Lousto, and Takahashi}}]{Baker:2002qf}
\bibinfo{author}{\bibfnamefont{J.}~\bibnamefont{Baker}},
  \bibinfo{author}{\bibfnamefont{M.}~\bibnamefont{Campanelli}},
  \bibinfo{author}{\bibfnamefont{C.~O.} \bibnamefont{Lousto}},
  \bibnamefont{and}
  \bibinfo{author}{\bibfnamefont{R.}~\bibnamefont{Takahashi}},
  \bibinfo{journal}{Phys. Rev. D} \textbf{\bibinfo{volume}{65}},
  \bibinfo{pages}{124012} (\bibinfo{year}{2002}{\natexlab{b}}),
  \eprint[http://arXiv.org/abs]{astro-ph/0202469}.

\bibitem[{\citenamefont{Boyle et~al.}(2007)\citenamefont{Boyle, Lindblom,
  Pfeiffer, Scheel, and Kidder}}]{Boyle:2006ne}
\bibinfo{author}{\bibfnamefont{M.}~\bibnamefont{Boyle}},
  \bibinfo{author}{\bibfnamefont{L.}~\bibnamefont{Lindblom}},
  \bibinfo{author}{\bibfnamefont{H.}~\bibnamefont{Pfeiffer}},
  \bibinfo{author}{\bibfnamefont{M.}~\bibnamefont{Scheel}}, \bibnamefont{and}
  \bibinfo{author}{\bibfnamefont{L.~E.} \bibnamefont{Kidder}},
  \bibinfo{journal}{Phys. Rev.} \textbf{\bibinfo{volume}{D75}},
  \bibinfo{pages}{024006} (\bibinfo{year}{2007}), \eprint{gr-qc/0609047}.

\bibitem[{\citenamefont{Tiglio et~al.}(2008)\citenamefont{Tiglio, Kidder, and
  Teukolsky}}]{Tiglio:2007jp}
\bibinfo{author}{\bibfnamefont{M.}~\bibnamefont{Tiglio}},
  \bibinfo{author}{\bibfnamefont{L.~E.} \bibnamefont{Kidder}},
  \bibnamefont{and} \bibinfo{author}{\bibfnamefont{S.~A.}
  \bibnamefont{Teukolsky}}, \bibinfo{journal}{Class. Quant. Grav.}
  \textbf{\bibinfo{volume}{25}}, \bibinfo{pages}{105022}
  (\bibinfo{year}{2008}), \eprint{0712.2472}.

\bibitem[{\citenamefont{Schnetter et~al.}(2006)\citenamefont{Schnetter, Diener,
  Dorband, and Tiglio}}]{Schnetter:2006pg}
\bibinfo{author}{\bibfnamefont{E.}~\bibnamefont{Schnetter}},
  \bibinfo{author}{\bibfnamefont{P.}~\bibnamefont{Diener}},
  \bibinfo{author}{\bibfnamefont{E.~N.} \bibnamefont{Dorband}},
  \bibnamefont{and} \bibinfo{author}{\bibfnamefont{M.}~\bibnamefont{Tiglio}},
  \bibinfo{journal}{Class. Quant. Grav.} \textbf{\bibinfo{volume}{23}},
  \bibinfo{pages}{S553} (\bibinfo{year}{2006}), \eprint{gr-qc/0602104}.

\bibitem[{\citenamefont{Zink et~al.}(2008)\citenamefont{Zink, Schnetter, and
  Tiglio}}]{Zink:2007xn}
\bibinfo{author}{\bibfnamefont{B.}~\bibnamefont{Zink}},
  \bibinfo{author}{\bibfnamefont{E.}~\bibnamefont{Schnetter}},
  \bibnamefont{and} \bibinfo{author}{\bibfnamefont{M.}~\bibnamefont{Tiglio}},
  \bibinfo{journal}{Phys. Rev.} \textbf{\bibinfo{volume}{D77}},
  \bibinfo{pages}{103015} (\bibinfo{year}{2008}), \eprint{0712.0353}.

\end{thebibliography}

\end{document}